\pdfoutput=1
\documentclass[11pt,a4paper]{article}
\usepackage{amsmath,amssymb}
\usepackage{graphicx}
\usepackage{cite}
\usepackage{url}

\renewcommand\({\left(}
\renewcommand\){\right)}
\renewcommand\[{\left[}

\def\beq{\begin{equation}}
\def\eeq{\end{equation}}

\begin{document}

\numberwithin{equation}{section}
\title{{\normalsize  \mbox{}\hfill DCPT/13/102, IPPP/13/51}\\
\vspace{2.5cm}
\Large{\textbf{Leptogenesis and Neutrino Oscillations in the Classically Conformal  Standard Model with the Higgs Portal}}}

\author{Valentin V. Khoze and Gunnar Ro\footnote{valya.khoze and g.o.i.ro@durham.ac.uk}\\[4ex]
  \small{\em Institute for Particle Physics Phenomenology, Department of Physics} \\
  \small{\em Durham University, Durham DH1 3LE, United Kingdom}\\[0.8ex]
}

\date{}
\maketitle

\begin{abstract}
  \noindent
The Standard Model with an added Higgs portal interaction and no explicit mass terms is a classically scale-invariant theory.
In this case the scale of electroweak symmetry breaking can be induced radiatively  by the Coleman-Weinberg mechanism 
 operational in a hidden sector, and then transmitted to the Standard Model through the Higgs portal.
The smallness of the generated values for the Higgs vev and mass, compared to the UV cutoff of our classically scale-invariant 
effective theory, is naturally explained by this mechanism.
 
We show how these classically conformal models can generate the baryon asymmetry of the Universe without the need of 
introducing mass scales by hand or their resonant fine-tuning. The minimal model we consider is the Standard Model coupled 
to the Coleman-Weinberg scalar field charged under the $U(1)_{B-L}$ gauge group. Anomaly cancellation requires automatic 
inclusion of three generations of right-handed neutrinos. Their GeV-scale Majorana masses are induced by the Coleman-Weinberg 
field and lead  to the generation of active neutrino masses through the standard see-saw mechanism. Leptogenesis occurs via 
flavour oscillations of right-handed sterile neutrinos and is converted to the baryon asymmetry by electroweak sphalerons. 
\end{abstract}
\thispagestyle{empty}
\setcounter{page}{0}

\newpage

\section{Introduction}\label{sec:intro}

Spontaneous breaking of the electroweak symmetry (EWSB) is one of the foundational concepts of contemporary particle physics.
That it occurs due to an elementary scalar Higgs field developing a non-vanishing vacuum expectation value, is
a fundamental principle put forward in the BEH mechanism~\cite{orig}, which is implemented in
the Standard Model (SM). 
The discovery of a $\simeq 125$ GeV scalar particle~\cite{ATLAS:2012gk,CMS:2012gu} with properties of the Higgs boson
is the crowning achievement of the SM as an effective theory.

At the more fundamental level, the SM leaves a number of key questions unanswered. First on the list is the question
of the origin of the electroweak scale and its naturalness. The SM accommodates the
expectation value $v= 246~{\rm GeV}$ for the Higgs field and the
Higgs mass $m_h \simeq 125~{\rm GeV}$ essentially as input parameters, but it does not (nor is it expected to) 
explain the origin or the value of the electroweak scale, and in particular its smallness 
compared to the UV cutoff or the scale of new physics. The list of questions left unanswered by the SM continues and
includes the generation of the baryon asymmetry of the Universe, the nature of Dark Matter and the particle physics 
implementation of inflation. In this paper we will argue that a minimal extension of the SM can naturally provide an
explanation for electroweak symmetry breaking and the baryon asymmetry. These two phenomena in fact will have the same
origin in the classically scale invariant theory.\footnote{An implementation of the cosmological inflation mechanism and the inclusion of the dark matter candidate in the classically scale-invariant 
extended Standard Model is considered in the follow-up paper.}

An elegant way to explain how the vev (and the mass) of the scalar field were generated and why their values are naturally suppressed
by many orders of magnitude relative to the UV scale at $\sim M_{\rm Pl}$,  
is to start from a purely massless theory, and to generate the mass gap radiatively.
In a seminal paper~\cite{Coleman:1973jx} Coleman and E. Weinberg showed
that in gauge theory a classically massless
\begin{equation}
  \label{eq:m0}
  m^2\, := \, \partial^2_{\phi}V(\phi)\bigg|_{\phi=0}=0 \, 
\end{equation}
scalar field $\phi$
does develop the vev
via dimensional transmutation
from the running couplings, leading to spontaneous  breaking of gauge symmetry. The vev is non-vanishing, calculable in a weakly-coupled theory
and is exponentially suppressed relative to the UV cutoff,
\begin{equation}
  \label{run2}
  \langle |\phi | \rangle \,  \sim\, M_{UV}\times \exp \left[ \frac{-24\pi^2}{g^{2}_{CW}(\langle |\phi |\rangle)}
  \right] \, \ll M_{UV}\,,
\end{equation}
where $g_{CW}$ is the gauge coupling of $\phi$, and for the vev we use\footnote{In what follows we 
do not distinguish between $\langle| \phi|^2 \rangle$,  $\langle |\phi | \rangle^2$ and $|\langle \phi \rangle|^2$.
}
 $\langle| \phi| \rangle:= \sqrt{\langle |\phi|^2\rangle}$.
\medskip

The use of the Coleman-Weinberg (CW) mechanism  is
based on approximate scale invariance of the theory. The scale invariance of the SM can either be seen as softly broken \cite{Bardeen:1995kv},
or it can be realised only as a classical symmetry.
The concept of
classical scale
invariance, first discussed in~\cite{Meissner:2006zh},
asserts that no masses are allowed in the classical Lagrangian of the effective theory 
under consideration; all mass scales must be generated dynamically in the IR. Even earlier, Ref.~\cite{Bardeen:1995kv}
considered the set-up where scale invariance is softly broken and discussed the appropriate choice of UV regularisation 
for such theories.

The minimal model to realise classical scale invariance is the Standard Model with an
additional CW ``hidden sector" and the Higgs portal-type coupling to
the SM~\cite{Hempfling:1996ht,Chang:2007ki,Foot:2007as,Iso:2009ss,Iso:2012jn,Englert:2013gz,Chun:2013soa,Heikinheimo:2013fta,Hambye:2013dgv}. The classical potential
for scalar fields in this minimal scenario is,
\begin{equation}
  \label{potentialcoupled1}
  V_{\rm cl}(H,\phi)\,=\, \frac{\lambda_{\rm H}}{2}(H^{\dagger} H)^{2}
   \,+\,\frac{\lambda_{\phi}}{4!}|\phi|^{4}
    \,-\, \lambda_{\rm P}(H^{\dagger}H)|\phi|^{2}\,.
\end{equation}
The first two terms are the ordinary self-couplings for
the Higgs field $H$ and the Coleman-Weinberg scalar $\phi$. The last term is the
Higgs-portal~\cite{Higgs.portal} interaction between the SM Higgs field and the hidden sector field
$\phi$. The portal coupling 
generates the negative Higgs-mass-squared  $=\,- \,\lambda_{\rm P}\,|\langle \phi\rangle|^{2}$ needed to trigger
EWSB at the scale $v$ which is exponentially suppressed relative to the UV cutoff scale as a consequence of \eqref{run2}.

To preserve classical scale invariance as an approximate quantum symmetry in the UV, only the scale-preserving 
UV regularisation schemes can be used~\cite{Bardeen:1995kv}. 
In dimensional regularisation, which does not introduce any explicit
scale aside from the RG scale, entering the logarithmically running
couplings, the masslessness equation \eqref{eq:m0} is satisfied automatically for all values of the RG scale
in theories which 
contain no explicit mass scales at the outset, and no {\it finite}
corrections to dimensionful quantities can appear either.

\bigskip

The Coleman-Weinberg field $\phi$ can be viewed as the pseudo-dilaton arising from the approximate scale invariance 
broken by the running of dimensionless couplings. In the classically scale-invariant BSM theory all mass scales should be linked
to and originate from $\langle \phi \rangle$. It is thus unlikely that a compelling explanation could be found 
for generation of vastly different scales in the theory. Thus it is a prediction of this general formalism for BSM model building with classical
scale invariance and at weak coupling that no vastly different scales can exist in the theory. The Planck scale can be taken separately
in this case, as gravity is viewed as a background to the effective classically scale-invariant theory. But all other
particle-physics scales, as soon as they are generated by dimensional transmutation through the vev of the Coleman-Weinberg 
pseudo-dilaton, should not be separated from one another by many orders of magnitude.

To give an example, this model-building principle
severely disfavours Grand Unification, since the GUT scale and the electroweak scale would have to have the same origin
in $\langle \phi \rangle$ and this would be quite difficult to implement given 14 orders of magnitude between them.

What about the baryon asymmetry?
With no signs of supersymmetry and no anomalies in the quark flavour sector, the most attractive (and arguably least unlikely) scenario 
for generating the baryon asymmetry of the Universe is leptogenesis. In the standard scenario of thermal leptogenesis \cite{Fukugita:1986hr}, a lepton asymmetry 
is generated by decays of heavy right-handed Majorana neutrinos into Standard Model leptons at temperatures much above the electroweak scale. 
The lepton asymmetry is then reprocessed into the baryon asymmetry by electroweak sphalerons 
\cite{Manton:1983nd, Kuzmin:1985mm}
above the electroweak scale. However, to generate the observed value of matter-anti-matter asymmetry in the vanilla version
of leptogenesis requires extremely heavy masses for sterile neutrinos, $M\gtrsim 10^{9}$ GeV \cite{Davidson:2002qv,Davidson:2008bu}.
If this was the full story, the classical scale-invariance would be ruled out by $M \ggg v$.
Instead we will adopt an alternative approach to leptogenesis pioneered in Ref.~\cite{Akhmedov:1998qx} 
and further developed in \cite{Asaka:2005pn,Drewes:2012ma}. In this approach, the lepton flavour asymmetry is being produced during 
oscillations of the right-handed Majorana neutrinos with masses of the order of the electroweak scale or below, which is perfectly suited for
our classically scale-invariant setup.

\bigskip
The paper is organised as follows.
In Section~{\bf \ref{sec:2.1}} we recall how the generation of the electroweak scale occurs in a minimal classically scale-invariant theory
based on the $U(1)_{CW}$ extension of the Standard Model. The $B-L$ realisation of this model which automatically includes 
sterile right-handed Majorana neutrinos is outlined in~{\bf \ref{sec:2.2}}. The formalism of leptogenesis via Majorana neutrino oscillations 
is presented in~{\bf \ref{sec:3.1}}. Section~{\bf \ref{sec:3.2}} adapts and applies theses ideas to our classically scale-invariant models.
The matter-anti-matter asymmetry is calculated and analysed in Section~{\bf \ref{sec:4}} which also contains our
benchmark points. Conclusions are outlined in Section~{\bf \ref{sec:five}}.

\bigskip
\section{EWSB in the classically conformal extension of the Standard Model}
\label{sec:2}

As we already noted, the classically scale-invariant extension of the Standard Model does not allow for
tree-level mass terms.  We now
recall how the Coleman-Weinberg mechanism leads to EWSB in the
Higgs-portal theory. Below we will summarise the relevant for us  
formulae following  \cite{Englert:2013gz}.

\subsection{The minimal $\mathbf{U(1)_{CW} \times SM}$ theory}
\label{sec:2.1}

The Coleman-Weinberg complex scalar
$\phi$ is coupled to a U(1)$_{CW}$ gauge theory
(this forms the hidden sector), while the Higgs doublet $H$ has
standard interactions with the SU(2)$\times$U(1) gauge fields (as well
as the matter fields) of the Standard Model.  
The classical scalar potential is given by\footnote{We use the
    normalisation of \cite{Coleman:1973jx} for the complex
    field $\phi=\phi_{1}+i\phi_{2}$ in terms of two real scalar fields with canonical kinetic terms
    $\frac{1}{2}(\partial_{\mu}\phi_{1}\partial^{\mu}\phi_{1}+
    \partial_{\mu}\phi_{2}\partial^{\mu}\phi_{2})$. This is related to the canonically normalised 
    complex scalar $S=(\phi_{1}+i\phi_{2})/\sqrt{2}$ via a simple rescaling,
    $S=\phi/\sqrt{2}$.}
\begin{equation}
  \label{potentialcoupled}
  V_{\rm cl}(H,\phi)\,=\, \frac{\lambda_{\rm H}}{2}(H^{\dagger} H)^{2}
  \,-\, \lambda_{\rm P}(H^{\dagger}H)|\phi|^{2}
  \,+\,\frac{\lambda_{\phi}}{4!}|\phi|^{4}\,,
\end{equation}
where
the Higgs doublet $H$ in the unitary gauge takes the form
$H^T(x)=\frac{1}{\sqrt{2}}(0, v+h(x))$ .

At the origin in field
space where all field vevs are zero, there are no scales
present in the theory.  This feature is maintained in the full effective
potential due to the renormalisation subtraction conditions,
\begin{equation}
  \label{rencond}
  \frac{\partial^{2} V(H,\phi)}{\partial H^{\dagger} \partial H}\bigg|_{H=\phi=0}\,=\, 0\,=\,
  \frac{\partial^{2} V(H,\phi)}{\partial \phi^{\dagger} \partial \phi}\bigg|_{H=\phi=0}\,.
\end{equation}
These conditions are automatic in
dimensional regularisation of any theory with classical scale
invariance and hold for any value of the RG scale $\mu$ in dimensional regularisation which does not break scale invariance.

In the phenomenologically relevant near-decoupling limit between the hidden and the SM sectors,
$\lambda_{\rm P}\ll 1$, we can  view
electroweak symmetry breaking effectively as a two-step process \cite{Englert:2013gz}.

First, the CW mechanism generates 
$\langle |\phi |\rangle$ in the CW sector as dictated by 
\eqref{run2}
through running of the gauge CW coupling.  At the scale $\mu=\langle |\phi |\rangle$ 
the scalar coupling $ \lambda_{\phi}$ is of the order of $g_{CW}^4 \ll 1$ rather than $g_{CW}^2$.

To see this, consider the 1-loop effective potential evaluated at the scale $\mu=\langle |\phi |\rangle$ ,
\begin{equation}
  \label{Veff2hp}
  V(\phi,H)\,=\, \frac{\lambda_{\phi}}{4!}|\phi|^{4}\,+\,
  \frac{3 g_{CW}^{4}}{64\pi^2}|\phi|^{4}
  \left[\log\left(\frac{|\phi|^{2}}{\langle |\phi|^{2}\rangle}\right)-\frac{25}{6}\right]
  \,-\, \lambda_{\rm P}(H^{\dagger}H)|\phi|^{2}\,+\,
  \frac{\lambda_{\rm H}}{2}(H^{\dagger} H)^{2}\,.
\end{equation}
Here following \cite{Englert:2013gz} we are keeping 1-loop corrections arising from interactions of
$\phi$ with the U(1) gauge bosons in the hidden sector, but neglecting
radiative corrections from the Standard Model sector. The latter would
produce only subleading corrections to the vevs.  The
$\phi$-minimisation condition
 gives 
\begin{equation}
  \label{Veffprime2}
  \partial_{\phi} V\,=\, \frac{1}{6}\left( \lambda_{\phi}-\frac{33}{8\pi^2}g^{4}_{CW}\right)
  {\langle |\phi|\rangle}^3   - 2\lambda_{\rm P}{\langle |H|^2 \rangle}{\langle |\phi| \rangle}\, =\,
  \frac{1}{6}\left( \lambda_{\phi}-\frac{33}{8\pi^2}g^{4}_{CW}-12\frac{\lambda_{\rm P}^2}{\lambda_{\rm H}}\right)
  {\langle |\phi| \rangle}^3  \, =\, 0\,.
\end{equation}
This equation implies that the vev $\langle |\phi |\rangle$ is determined by the
condition on the four couplings renormalised at the scale of the vev
\begin{equation}
  \label{eq:rad4}
  \lambda_{\phi}(\langle|\phi|\rangle)-\frac{33}{8\pi^2}g^{4}_{CW}(\langle |\phi|\rangle) \, =\, 
  12\frac{\lambda_{\rm P}^2 (\langle |\phi|\rangle)}{\lambda_{\rm H}(\langle |\phi|\rangle)} \,\simeq \,0 \,.
\end{equation}
For small $\lambda_{\rm P}$, this is a small deformation of the
original CW condition, $\lambda_{\phi}(\langle|\phi|\rangle) =\, \frac{33}{8\pi^2}g^{4}_{CW}(\langle |\phi|\rangle)$.

\medskip

The second step of the process is the transmission of the vev $\langle |\phi | \rangle$  to the
Standard Model via the Higgs portal, generating a negative mass squared 
parameter for the Higgs $=-\lambda_{\rm P}\langle |\phi|^{2}\rangle$
which fixes the
electroweak scale:
\begin{equation}
  \label{musm3}
 \frac{1}{2\lambda_{\rm P}} \, 
  (125~{\rm GeV})^{2}  \, =\, 
  \frac{\lambda_{\rm H}}{2\lambda_{\rm P}} \, {(246~{\rm GeV})^{2}} \, =\, 
   \langle |\phi|^{2}\rangle \,,
\end{equation}
where we used $m_{h} = 125~{\rm GeV}$ and $\sqrt{2} \langle H \rangle = 246~{\rm GeV}$.
The fact that for $\lambda_{\rm P}\ll 1$ the generated electroweak scale is much smaller than
$\langle  |\phi | \rangle$, guarantees that any back reaction on the hidden
sector vev $\langle |\phi | \rangle$ is negligible.

The two vevs, $\langle |\phi |\rangle$ and $v$ are generated {\it
  naturally} through dimensional transmutation in our framework
similarly to \eqref{run2},
\begin{equation}
  \label{run3}
  \sqrt{\frac{\lambda_{\rm H}}{2\lambda_{\rm P}}} \,v \,=\,
  \langle |\phi |\rangle  \, \simeq \, \Lambda_{UV}\exp 
  \left[ \frac{-24\pi^2}{g^{2}_{CW}(\langle |\phi|\rangle)}
  \right] \, \ll \Lambda_{UV}\,.
\end{equation}

In the hidden sector we have two additional fields: $\phi$ and the
extra U(1)$_{\rm hidden}$ gauge field $Z'$.  After $\phi$ acquires a
non-vanishing vev, the $Z'$ becomes massive\footnote{If the U(1)$_{CW}$
charge $Q_{\phi}$ of the CW scalar field is different from unity, as will be the case for the $B-L$ model 
discussed in the following section, the gauge coupling in the expressions on the right hand side of Eq.~\eqref{eqn:MZ} and in the second 
equation in \eqref{hid-scs} should be rescaled as $g_{CW} \rightarrow Q_\phi\, g_{CW}$.}
\begin{equation}
\label{eqn:MZ}
m_{Z'}=g_{CW}\langle |\phi |\rangle\,.
\end{equation}
The  $Z'$ boson is much heavier than the CW scalar, since from \eqref{Veff2hp} we have
$m_{\varphi}^2 = \frac{3 g^2_{CW}}{8\pi^2} m_{Z'}^2$,
thus $m_{Z'}$ would have to be in the few TeV range in order to be not yet seen at the LHC.

Let us count the parameters of the model:  the SM Higgs
self-coupling is fixed by the ratio of known electroweak scales, while
the other self-coupling, $\lambda_{\phi}$, is determined from the CW
dimensional transmutation condition \eqref{eq:rad4}. 
There are two undetermined parameters in our model which one can take
to be the hidden sector gauge coupling, $g^2_{CW}$, and the (small)
portal coupling $\lambda_{\rm P}$. In this case, the two mass scales
associated with the hidden scalar are fixed,
\begin{equation}
  \label{hid-scs}
  \langle |\phi|^2 \rangle \,=\, \frac{1}{2 \lambda_{\rm P}}\, m_h^2
  ,\qquad
  m_{\varphi}^2 \,=\,\frac{3 g^4_{CW}}{8\pi^2}\,\langle |\phi|^2
  \rangle \,=\,
  \frac{3 g^4_{CW}}{16\pi^2}\,\frac{1}{\lambda_{\rm P}}\, m_h^2\,.
\end{equation}
Alternatively, the two free parameters can be chosen to be the mass of
the hidden Higgs, $m_{\varphi}$, and the Higgs portal coupling
$\lambda_{\rm P}$. 

\bigskip

The
phenomenology of this model in the context of LHC, future
colliders and low energy measurements was analysed in \cite{Englert:2013gz}.
The minimal model has only two
remaining free parameters, the mass of the CW scalar\footnote{More precisely, since the Higgs field and the CW field mix, it is the mass of the second 
scalar eigenstate, with the first being the $m_h=125$ GeV Higgs. This parameter can also be traded for the mass of the CW gauge field $Z'$.}, and the portal coupling $\lambda_{\rm P}$,
and it was shown 
that the model is perfectly
viable. In particular, the presently available Higgs data 
constrains the portal coupling to be $\lambda_{\rm P} \lesssim 10^{-5}$ on the `half' of the parameter space where the second scalar is in the region between $10^{-4}$ GeV and 
$m_h/2$. 

At the same time, inside a much more restricted  window in the parameter space, where the second scalar is heavier than $m_h/2$, but less than 100 GeV, the portal coupling can be pushed up to the
$\lambda_{\rm P} \sim 10^{-3}$ regime. In this narrow window, the Higgs potential is automatically stabilised at high scales by the positive contribution 
$\propto \lambda_{\rm P}^2$ 
to the Higgs self-coupling $\lambda_H$ beta function, a recent discussion of this effect can be found in  \cite{Hambye:2013dgv, Lykken}, see also \cite{EliasMiro:2012ay}.
As the result, here the EWSB vacuum at $v=246$ GeV is stable at all energy scales.

When the second scalar is heavier than 100 GeV, the model is largely unconstrained by current experimental data, see Ref.~\cite{Englert:2013gz}
for more detail.
Future experimental data on
Higgs decays will further constrain model parameters, and will ultimately provide discovery potential for this model.

\subsection{The $\mathbf{B-L}$ Coleman-Weinberg extension of the Standard Model}
\label{sec:2.2}

The Coleman-Weinberg mechanism must occur in the sector separate from the Standard Model.
This is because 
 the CW scalar must be parametrically lighter than the vector boson of the gauge group under which the CW scalar is charged,
{\it cf.} the second equation in \eqref{hid-scs},
\begin{equation}
  \label{hid-scs2}
    m_{\varphi}^2 \,=\,\frac{3 }{8\pi^2}\,(Q_\phi\,g_{CW})^2 m_{Z'}^2 \ll \,m_{Z'}^2\,.
\end{equation}
Hence the CW gauge boson is identified with $Z'$ rather than with the $Z$ boson of the SM $SU(2)_L$ which is of course lighter than its scalar (i.e. the Higgs).
The minimal CW gauge group is $U(1)$ (though nothing prevents the CW mechanism to work well in the non-Abelian settings \cite{Coleman:1973jx};
for a recent $SU(2)$ CW application see  \cite{Hambye:2013dgv}).

In Eq.~\eqref{hid-scs2} we have allowed for a general $Q_\phi$, which denotes the charge of the CW scalar under the U(1) gauge group; it will be set to $Q_\phi=2$ in the $B-L$ model we are going to consider.

\bigskip
An interesting idea~\cite{Iso:2009ss,Iso:2012jn} is to identify the CW $U(1)$ factor with the gauged $B-L$ $U(1)$ flavour subgroup of the Standard Model.
This gives the classically conformal realisation of what is known as the $B-L$ model~\cite{Mohapatra:1980qe,Basso:2008iv}. 

We thus have $g_{CW}\equiv g_{B-L}$, and the $Z'$ massive ($\gtrsim$ few TeV)
vector boson now couples to quarks and leptons of the Standard Model
proportionally to their ${B-L}$ charge. The SM Higgs carries no baryon or lepton number and therefore does not couple to the $U(1)_{B-L}$ sector. 

The appeal of this model with local $U(1)_{B-L}$ group is that the cancellation of gauge anomalies requires an automatic inclusion of 
three generations of the right-handed neutrinos, ${\nu_{R}^{}}_i$. These neutrinos carry lepton number = 1 and transform under $U(1)_{B-L}$,
but are sterile under the SM gauge groups. Finally, the Coleman-Weinberg scalar field $\phi$ is assigned the $B-L$ charge $=2$.
Interactions of right-handed neutrinos ${\nu_{R}^{}}_i$ are given by
\begin{equation}
  \label{eq:Lnu}
 {\cal L}^{\nu_R}_{\rm int}\,=\, - \frac{1}{2}\( Y^{\rm M}_{ij} \phi\, \overline{\nu_{R}^c}_i {\nu_{R}^{}}_j \,+\, 
 Y^{\rm M \, \dagger}_{ij} \phi^{\dagger}\, \overline{\nu_{R}^{}}_i {\nu_{R}^{c}}_j  \) 
 \,-\, Y^{\rm D}_{ia} \overline{\nu_{R}^{}}_i (\epsilon H)\, l^{}_{L\,a}
 \,-\, Y^{\rm D \,\dagger}_{ai} \,\overline{l_{L}^{}}_a (\epsilon H)^\dagger \, \nu^{}_{R\,i}
\end{equation} 
where $Y^{\rm M}_{ij}$ and $Y^{\rm D}_{ia} $ are $3 \times 3$ complex matrices of the Majorana and Dirac Yukawa couplings respectively.
The right-handed neutrinos ${\nu_{R}^{}}_i$ are SM singlets (often referred to as the sterile neutrinos), they carry lepton number $L=+1$ and their antiparticles, $\overline{\nu_{R}^{}}_i$, have $L=-1$. The charge-conjugate anti-particle,
$\overline{\nu_{R}^c}_i $ has the same lepton number +1 as the state ${\nu_{R}^{}}_i$. In the unbroken phase, the lepton number is conserved by all interactions in \eqref{eq:Lnu} when $\phi$ is assigned lepton number $-2$. The first two terms on the right hand side of \eqref{eq:Lnu} 
give the only interactions of the CW scalar $\phi$ with matter fields (apart from its small mixing with the Higgs). This is a consequence of its
$L=-2$ charge assignment.

Spontaneous breaking of the $B-L$ symmetry by the vev 
$\langle |\phi |\rangle \neq 0$ generates Majorana masses 
\begin{equation}
  \label{eq:Maj}
 M_{ij} \,=\, Y^{\rm M}_{ij} \, \langle |\phi |\rangle
 \end{equation}  
which lead to lepton number non-conserving interactions. Importantly, individual lepton flavour is also not conserved: $M_{ij}$ are complex matrices which induce CP-violating transitions between lepton flavours $i$ and $j$ of the right-handed neutrinos.\footnote{This is most easily seen in the ``Dirac-Yukawa basis'' where the Dirac Yukawa matrices 
$Y^{\rm D}_{ia} $ are diagonalised and real, but not the Majorana ones $Y^{\rm M}_{ij}$.}

In summary, the single $U(1)_{B-L}$ hidden sector simultaneously incorporates the Coleman-Weinberg scalar which triggers the EWSB, and 
also gives rise to
the Majorana sterile neutrinos which through the see-saw mechanism give rise to masses of active neutrinos and neutrino oscillations\cite{Iso:2009ss,Iso:2012jn}.
Furthermore, as will be shown below, the generation of matter-antimatter asymmetry through leptogenesis now becomes possible
and without fine-tuning.

\bigskip
\section{Neutrino oscillations and leptogenesis}
\label{sec:3}

Leptogenesis is the idea that the baryon asymmetry of the Universe has originated in the lepton rather than quark sector of the theory. In the standard scenario of thermal leptogenesis  \cite{Fukugita:1986hr} one starts with the see-saw Lagrangian involving right-handed neutrinos with Majorana mass terms coupled to the Standard Model left-handed lepton doublets
({\it cf.} Eq.~\eqref{eq:Lnu}),
\begin{equation}
  \label{eq:LnuM}
 {\cal L}^{\nu_R}_{\rm int}\,=\, - \frac{1}{2}\( M_{ij}\, \overline{\nu_{R}^c}_i {\nu_{R}^{}}_j \,+\, 
 M_{ij}^{\dagger}\, \overline{\nu_{R}^{}}_i {\nu_{R}^{c}}_j  \) 
 \,-\, Y^{\rm D}_{ia} \overline{\nu_{R}^{}}_i (\epsilon H)\, l^{}_{L\,a}
 \,-\, Y^{\rm D \,\dagger}_{ai} \,\overline{l_{L}^{}}_a (\epsilon H)^\dagger \, \nu^{}_{R\,i}\,.
\end{equation} 
It is usually assumed that a lepton asymmetry 
was generated by decays of heavy right-handed Majorana neutrinos at temperatures much above the electroweak scale. These heavy sterile neutrinos were thermally produced during reheating in the early Universe and then fell out of thermal equilibrium due to the Universe expansion. 
Their out-of-equilibrium decays into Standard Model leptons and Higgs bosons violate lepton number and CP, thus producing lepton asymmetry, which is then reprocessed into the baryon asymmetry by electroweak sphalerons above the electroweak scale. 

The defining phenomenological signature of these models is that the masses of the sterile Majorana neutrinos should be 
$M\gtrsim 10^{9}$ GeV \cite{Davidson:2002qv,Davidson:2008bu}.
Flavour effects~\cite{Barbieri:1999ma} and a resonant enhancement~\cite{Pilaftsis:2003gt}
are important and can somewhat lower this bound, but not by many orders of magnitude\footnote{Unless one is willing to fine-tune sterile neutrino masses of different flavours to introduce mass degeneracy $M_i M_j/|M^2_i-M^2_j| \ggg 1$. 
This is not the approach we will follow.}.

\subsection{Leptogenesis triggered by oscillations of Majorana neutrinos}
\label{sec:3.1}

Akhmedov, Rubakov and Smirnov (ARS) in Ref.~ \cite{Akhmedov:1998qx} proposed an alternative physical realisation of the leptogenesis mechanism which allows one to circumvent the $\sim 10^{9}$ GeV lower bound.
In fact, the ARS leptogenesis is intended to work with sterile neutrinos of sub-electroweak Majorana mass-scale. The generation of matter-anti-matter asymmetry proceeds as follows. As in the original mechanism, the right-handed neutrinos 
are produced thermally in the early Universe through their Yukawa interactions with lepton and Higgs doublets. 
After being produced, they begin to oscillate,   ${\nu_{R}^{}}_i \, \leftrightarrow\, {\nu_{R}^{}}_j$, between the three different flavour states
$i,j=1,2,3$ in the expanding Universe, and also interact with the left-handed leptons and Higgs bosons via their Yukawa interactions.

Since the Majorana masses in the ARS scenario are roughly of the electroweak scale, or below, they are much smaller than the relevant temperature, $T_{\rm osc}$, in the early Universe. For this reason, the rate of the total lepton-number violation (i.e. singlet fermions to singlet anti-fermions, ${\nu_{R}^{}}_i \, \leftrightarrow\, \overline{\nu_{R}^{}}_j$, induced by their 
Majorana masses) is negligible at $M/T_{\rm osc} \ll 1$. However, the lepton number of individual flavours is not conserved: 
complex non-diagonal Majorana matrices induce CP-violating flavour oscillations followed by out-of-equilibrium
-- due to smallness of the Yukawa matrices at $T_{\rm osc}$ -- decays,
\begin{equation}
  \label{eq:ARSflav}
{\nu_{R}^{}}_i \, \leftrightarrow\, {\nu_{R}^{}}_j\, \to\,{ l_{L}^{}}_j \, H 
\end{equation}
Following \cite{Akhmedov:1998qx} we now require that by the time the temperature cools down to $T_{EW}$, where 
electroweak sphaleron processes freeze out, two of the neutrino flavours i.e. $ {\nu_{R}^{}}_2$ and
$ {\nu_{R}^{}}_3$ equilibrate 
with their Standard Model counterparts,  ${ l_{L}^{}}_{2,3} \, H $, while the remaining flavour (call it the 1$^{\rm st}$ or $e$-flavour) 
does not.\footnote{The opposite case where only one flavour equilibrates before the sphaleron freeze-out, can be treated similarly. 
Essentially, {\it both} cases can be treated by not imposing any constraint on $\Gamma_2$, i.e. by simply dropping the first equation in
\eqref{eq:ARSequil}.}
In terms of the decay rates for the three sterile neutrino flavours this implies,
\begin{equation}
  \label{eq:ARSequil}
 \Gamma_2(T_{EW}) \,>\, H(T_{EW}) \ , \quad 
  \Gamma_3(T_{EW}) \,>\, H(T_{EW}) \ , \quad \Gamma_1(T_{EW}) \,<\, H(T_{EW}) 
\end{equation}  
where $H$ is the expansion rate of the Universe given by the Hubble `constant' 
\begin{equation}
  \label{eq:H}
 H(T) \,=\, \frac{T^2}{M_{\rm Pl}^*} \, , \qquad 
 M_{\rm Pl}^*\,\equiv \, \frac{M_{\rm Pl}}{\sqrt{g_*}\sqrt{4\pi^3/45}} \,\simeq\, 10^{18}\, {\rm GeV}
  \end{equation} 
 and $M_{\rm Pl}^*$ is the reduced Planck mass.
 
 As the result of this washout of the second and the third lepton flavours, the corresponding lepton doublets are processed by electroweak
 sphalerons into baryons, while the first flavour of right-handed neutrinos is not transferred to the active leptons fast enough before the electroweak sphaleron shuts down. (If the sphaleron did not freeze out below $T_{EW}$, all three flavours would have had enough time to thermalise and the net lepton and baryon asymmetry would have been zero.)


In the ARS approach\footnote{Reader primarily interested in the final expression for the lepton asymmetry can skip directly
to Eqs.~\eqref{eq:dzDG}-\eqref{eq:deltaLDG} which summarise the main result as derived in Ref.~\cite{Drewes:2012ma}. }
the interactions of sterile Majorana neutrinos ${\nu_{R}^{}}_i$ with the thermal plasma are
described by the $3 \times 3$ density matrix $\rho_{ij}$ with the evolution equation \cite{Sigl:1992fn}
\begin{equation}
  \label{eq:ARSrho}
 i \frac{d \rho}{dt}\,=\,  [{\cal H},\rho] - \frac{i}{2} \{\Gamma, \rho\} + i\Gamma^{\rm p}\, ,
\end{equation} 
where ${\cal H}$ is the Hermitian effective Hamiltonian, and $\Gamma$ and $\Gamma^{\rm p}$ are the destruction and production rates of ${\nu_{R}^{}}_i$. In the Yukawa basis at temperatures much higher than the Majorana mass, the effective Hamiltonian is of the form 
\begin{equation}
  \label{eq:Ham}
{\cal H}\,=\, U\, \frac{\hat{M}^2}{2 k(T)}\,U^{\dagger}\,+\, V(t)
 \end{equation}
where $U$ is the mixing matrix which relates the Yukawa basis with the mass eigenstate basis where the Majorana masses are diagonal, 
$\hat{M}^2 = {\rm diag} (\hat{M}_1^2\,,\,\hat{M}_2^2\,,\,\hat{M}_3^2)$, and $k(t)\simeq T$ is the neutrino momentum.

The first term on the right hand side of \eqref{eq:Ham} is the free Hamiltonian describing sterile neutrino oscillations
-- it originates from a tree-level diagram of ${\nu_{R}^{}}_i$ to ${\nu_{R}^{}}_j$ propagation 
with two helicity flips $\propto (M/2)^2$ connected by the propagator $2/k(T)$.
The second term in \eqref{eq:Ham} is the potential due to coherent forward scattering processes \cite{Akhmedov:1998qx},
\begin{equation}
  \label{eq:V}
V\,=\, {\rm diag} \(V_1,V_2,V_3\)    \, , \qquad
V_i\,=\, \frac{1}{8} (Y^{\rm D}_i)^2\, T
 \end{equation}

For the destruction rates of the sterile neutrino in \eqref{eq:ARSrho} ARS take the dominant Higgs-mediated two-to-two processes involving a lepton and a top-anti-top pair,
\begin{equation}
  \label{eq:Gamma}
\Gamma\,=\, {\rm diag} \(\Gamma_1,\Gamma_2,\Gamma_3\)    \, , \qquad
\Gamma_i\,\sim \, \frac{9 y_t^2}{64\pi^3} (Y^{\rm D}_i)^2\, T
 \end{equation}
where $y_t$ is the top Yukawa. More accurately, the destruction (or relaxation) rates of sterile neutrinos can
be accounted for as follows \cite{Drewes:2012ma},
\begin{equation}
  \label{eq:GammaDG}
\Gamma_i\,=\, \sum_a
Y^{\rm D }_{ia} Y^{\rm D \,\dagger}_{ai} \,\gamma_{av} \, T
 \end{equation}
 Here $\gamma_{av} $ is the dimensionless quantity inferred from the rates 
 tabulated in Ref.~\cite{Besak:2012qm}, it has a weak
 dependence on temperature, so that at $T= 5 \times 10^5$ GeV,   $\gamma_{av} \simeq 3 \times 10^{-3}$ while
 at electroweak temperature, $\gamma_{av} (T_{EW})\, \simeq 5 \times 10^{-3}$. Below, following  \cite{Drewes:2012ma},
 we will use 
 \eqref{eq:GammaDG} for the relaxation rate (we also note that this expression is  written in the basis-independent form).
 
 The final ingredient appearing in the ARS kinetic equation \eqref{eq:ARSrho} is the production rate $\Gamma^{\rm p}$
 which is determined in terms of the destruction rate $\Gamma$ above and the equilibrium density matrix, 
 $i \Gamma^{\rm p} \,=\, i \Gamma \rho^{\rm eq} \,=\, i \exp (-{\rm k}/T)\, \Gamma$, \cite{Akhmedov:1998qx}.
 
The production of the asymmetry starts at the time $t_{\rm osc}$ which corresponds to the temperature $T_{\rm osc}$
when the sterile neutrinos have performed at least 
one oscillation. This happens when the difference of the eigenvalues of the free Hamiltonian in \eqref{eq:Ham}
becomes of the order of the Hubble constant \cite{Akhmedov:1998qx},
\begin{equation}
  \label{eq:Tosc}
 \frac{|M_{i}^2-M_{j}^2|}{2 T_{\rm osc}} \,=\, 2\pi \, H(T_{\rm osc}) \quad => \quad
 T_{\rm osc}\,=\, \left(\frac{|M_{i}^2-M_{j}^2| M_{\rm Pl}^*}{4\pi}\right)^{1/3}
\end{equation}

Lepton flavour asymmetry is converted by electroweak sphalerons to baryon asymmetry
until the process ends at $T_{EW} \simeq 140$ GeV, when the sphalerons freeze-out. By this time the
equilibration of {\it two out of three} neutrino flavours occurs as in Eq.~\eqref{eq:ARSequil}, which is required for a non-vanishing
amount of the asymmetry generated.
The equilibration condition (i.e. wash-out) `boundary' determines the relevant values of the Yukawas via,
\begin{equation}
  \label{eq:equil}
\sum_a
Y^{\rm D }_{ia} Y^{\rm D \,\dagger}_{ai} \,\frac{\gamma_{av} \, T_{EW}}{H(T_{EW})}\,\sim\, 1
\quad =>
\quad
\sum_a
Y^{\rm D }_{ia} Y^{\rm D \,\dagger}_{ai} \,\simeq\, \frac{2.0}{\gamma_{\rm av}} \times 10^{-16} \,\simeq\,
4 \times 10^{-14}
 \end{equation}

Now, by integrating the kinetic equation for the sterile neutrino density matrix \eqref{eq:ARSrho} between
$t_{\rm osc}$ to $t_{EW}$,
the authors of \cite{Akhmedov:1998qx} were able to derive the expression for the number density $n_1 := \rho_{11}$ of the
lepton flavour asymmetry carried by the unequillibrated flavour. Up to an overall numerical factor and combining the neutrino mixing matrix angles together with the CP phase $\delta$ into a Jarlskog invariant 
$J=\, s_{12} c_{12}s_{13} c_{13}^2 s_{23} c_{23} \,\sin \delta$, the functional form of the generated lepton asymmetry over the entropy density of the Universe $s$ reads schematically\footnote{We will write down the precise and improved expression 
in Eq.~\eqref{eq:deltaLDG} below.}
 \begin{equation}
  \label{eq:deltaLARS} 
  {\rm ARS:} \qquad
\frac{n_L}{s} \,\sim\, J \, \,\frac{\Delta(Y^D)^2 \Delta(Y^D)^2 \Delta(Y^D)^2\, 
(M_{\rm Pl}^*)^2}{|\Delta M^2|^{1/3} |\Delta M^2|^{1/3} |\Delta M^2|^{1/3}}\,\, \gamma_{av} 
\end{equation}

Seven years after ARS, in Ref.~\cite{Asaka:2005pn},   Asaka and Shaposhnikov (AS) extended this approach by including
the back-reaction of active neutrinos on the sterile neutrinos. Specifically, the authors of \cite{Asaka:2005pn} 
have solved the kinetic equation \eqref{eq:ARSrho} for the $12 \times 12$ density matrix
whose components describe the mixing of all active and sterile neutrinos and anti-neutrinos,
 \begin{equation}
  \label{eq:rhoAS} 
  \rho\,=\,
 \left( \begin{matrix}
  \rho_{i\,j} &\rho_{i \,\bar{j}} & \rho_{i\,b} & \rho_{i\,\bar{b}}\\
  \rho_{\bar{i}\,j} & \rho_{\bar{i}\,\bar{j}} & \rho_{\bar{i}\,b} & \rho_{\bar{i}\,\bar{b}}\\
  \rho_{a\,j} &\rho_{a \,\bar{j}} & \rho_{a\,b} & \rho_{a\,\bar{b}}\\
  \rho_{\bar{a}\,j} & \rho_{\bar{a}\,\bar{j}} & \rho_{\bar{a}\,b} & \rho_{\bar{a}\,\bar{b}}
  \end{matrix}\right) \, \simeq \,
  \left( \begin{matrix}
  \rho_{i\,j} &0 & 0 & 0\\
 0 & \rho_{\bar{i}\,\bar{j}} &0& 0\\
  0 & 0 & \rho_{a\,b} & 0\\
  0&0&0& \rho_{\bar{a}\,\bar{b}}
  \end{matrix}\right)
 \end{equation} 
Here the elements of the density matrix  which mix sterile with active (anti)-neutrinos are neglected as they describe correlations
between particles of very different masses. Also the elements mixing neutrinos with anti-neutrinos are dropped as they give lepton number (or helicity) flips.
The resulting $\rho$-matrix is an extension of the simple $3 \times 3$ sterile-to-sterile density matrix 
$\rho_{i\,j}$ (and its CP-conjugate $\rho_{\bar{i}\,\bar{j}}$) used by ARS, as reviewed above.

The functional form of the generated lepton asymmetry computed by AS is given by ({\it cf.} Eq.~\eqref{eq:deltaLARS}),
\begin{equation}
  \label{eq:deltaLAS} 
  {\rm AS:} \qquad
\frac{n_L}{s} \,\sim\, \tilde{J } \, \,\frac{Y^D\, Y^D\, \Delta(Y^D)^2\, 
(M_{\rm Pl}^*)^{4/3}}{|\Delta M^2|^{1/3} |\Delta M^2|^{1/3} }\,\, \gamma_{av} ^2\, ,
\end{equation}
where $\tilde{J }$ is a certain combination of mixing angles and CP phases.

Quite remarkably, the functional form of the Asaka-Shaposhnikov result in \eqref{eq:deltaLAS} was fully reproduced by the
recent more technical derivation of the lepton asymmetry by Drewes and Garbrecht (DG) in \cite{Drewes:2012ma}. 
Their approach is based on a systematic application of non-equilibrium QFT methods (the Schwinger-Keldysh formalism
\cite{Schwinger:1960qe,Keldysh:1964ud}) to the calculation of the lepton flavour asymmetry, see also 
\cite{Garbrecht:2011aw,Garny:2011hg}. It is this result of \cite{Drewes:2012ma}  (which in the following Section will be adopted to
the case of the Coleman-Weinberg $B-L$ model with the $\langle \phi \rangle$-induced and thermally corrected Majorana masses) 
which we will use for our calculation of the resulting matter-anti-matter asymmetry.

Having noted the  fact that the non-equilibrium calculation of \cite{Drewes:2012ma} reproduces the parametric form (though with a different numerical factor)
of the more intuitive formalism of AS based on the density matrix, we can now proceed to simply state
the equation which determines the generation of lepton asymmetry in \cite{Drewes:2012ma},
\begin{equation}
  \label{eq:dzDG} 
 \frac{d}{dz} \frac{n_{La}}{s}\,=\, \frac{2\, S_{aa}}{s T_{EW}}  \, ,
 \end{equation}
 where 
 $n_{La}$ is the produced charge density of active lepton number of flavour $a$ (particles minus anti-particles),
 $s=\frac{2\pi^2}{45}g_* T^3$ is the entropy density of the Universe and 
 the `time' variable $z$ is defined via $z:= T_{EW}/T$. On the right hand side we have the source term given by the
 expression \cite{Drewes:2012ma},
 \begin{equation}
  \label{eq:DGsource} 
 \frac{2\, S_{aa}}{s T_{EW}} \,=\, - \sum_{c}\sum_{i\neq j} i \, 
 \frac{Y^{\rm D \,\dagger}_{ai} Y^{\rm D}_{ic} Y^{\rm D \,\dagger}_{cj} Y^{\rm D }_{ja} -
 Y^{\rm D \,t}_{ai} Y^{\rm D\, *}_{ic} Y^{\rm D \,t}_{cj} Y^{\rm D \, *}_{ja}}{M_{ii}^2-M_{jj}^2}\,
 \frac{M_{\rm Pl} T_{EW}}{z^2}\, \gamma_{av} ^2 \times 7.3 \times 10^{-4}\, .
 \end{equation}
 To determine the lepton asymmetry we integrate $\int_{z_{\rm osc}}^1 2\, S_{aa} /(s T_{EW}) \,dz$ using
 the expression in \eqref{eq:DGsource}.
 The lower limit
 $ z_{\rm osc}$ corresponds to the early temperature $T_{\rm osc}$ in \eqref{eq:Tosc}
 where the oscillations of sterile neutrinos start competing with the Hubble rate,
\begin{equation}
  \label{eq:zosc} 
 z_{\rm osc}^3 \,:=\,\left( \frac{T_{EW}}{T_{\rm osc}}\right)^3 \,=\, 8\pi\, \sqrt{\frac{\pi^3 g_* }{45}}
 \frac{T_{EW}^3}{M_{\rm Pl} |M_{ii}^2-M_{jj}^2| }\, .
 \end{equation}
The upper integration limit $z=1$ is the electroweak phase transition temperature, $T_{EW}$ where
the sphaleron freezes out. 
The integral gives the desired lepton asymmetry, which is the main result of \cite{Drewes:2012ma},
\begin{equation}
  \label{eq:deltaLDG} 
  {\rm DG:} \quad
\frac{n_{La}}{s} \,=\, - \sum_{c}\sum_{i\neq j} i \, 
 \frac{Y^{\rm D \,\dagger}_{ai} Y^{\rm D}_{ic} Y^{\rm D \,\dagger}_{cj} Y^{\rm D }_{ja} -
 Y^{\rm D \,t}_{ai} Y^{\rm D\, *}_{ic} Y^{\rm D \,t}_{cj} Y^{\rm D \, *}_{ja}}{{\rm sign}(M_{ii}^2-M_{jj}^2)}
\left( \frac{M_{\rm Pl}^2}{ |M_{ii}^2-M_{jj}^2|}\right)^{\frac{2}{3}} 
\gamma_{av} ^2 \times 1.2 \times 10^{-4}\,.
\end{equation}

\subsection{Leptogenesis in classically massless models }
\label{sec:3.2}

The focus of this paper are BSM models with classical scale invariance. In these
models no explicit mass scales are allowed in the Lagrangian as they would break classical scale invariance, hence 
all masses have to be generated dynamically, e.g. by vacuum expectation values 
of scalars induced by the Coleman-Weinberg field. 

In the minimal $B-L$ model, Majorana masses $M_{ij}$ for right-handed neutrinos 
are generated by the vev $\langle \phi \rangle$
of the Coleman-Weinberg field\footnote{In more general
settings, the sterile neutrinos could couple to a different scalar which would get its vev through a portal coupling to the
Coleman-Weinberg field. In this paper we concentrate on the minimal case where the scalar
responsible for the Majorana mass of sterile neutrinos is the Coleman-Weinberg field itself. Extensions with more scalars are straightforward. }
in Eq.~\eqref{eq:Maj}.
There are two effects which need to be taken into account. One is that at temperatures above the critical temperature 
$T_{B-L} \sim \langle |\phi |\rangle$, the spontaneously broken  $U(1)_{B-L}$ gauge symmetry is restored, so that in the
unbroken phase the Coleman-Weinberg field vev vanishes, $\langle |\phi |\rangle =0$. Secondly, due to interactions
of right-handed neutrinos with $\phi$ and with the $B-L$ gauge bosons, $Z'$, there
are also thermal corrections which need to be taken into account. 

To do this we write down the effective Hamiltonian \eqref{eq:Ham}  in the form
\begin{equation}
  \label{eq:HamT}
{\cal H}\,=\, \frac{M^2}{2 T}\,+\, V^{\rm M}(T) \,+\, V^{\rm D}(T) \,, 
 \end{equation}
where the first term is the tree-level effect of Majorana mass insertions as before, it is now given by
\begin{equation}
 \label{eq:HamT1}
\frac{M^2}{2 T} \,=\, \frac{|Y^{\rm M}|^2_{ij} \, |\langle \phi \rangle|^2}{2T}\,\, \Theta(T_{B-L}-T)\,\simeq\,
\frac{|Y^{\rm M}|^2_{ij} \, |\langle \phi \rangle|^2}{2T}\,\, \Theta(\langle |\phi |\rangle-T)\,.
 \end{equation}
Here the theta-function accounts for the transition to the unbroken phase at temperatures above $T_{B-L} \sim \langle |\phi |\rangle$.

The second term on the right hand side of \eqref{eq:HamT}  takes into account new self-energy diagrams for the right-handed neutrino due to interactions with the Coleman-Weinberg scalar $\phi$ and the $Z'$ bosons,
\begin{equation}
  \label{eq:HamVM}
V^{\rm M}\,=\, \frac{1}{32} |Y^{\rm M}|^2_{ij}\, T\,+\,\frac{1}{8} g_{B-L}^2 \,\delta_{ij}\,T
 \end{equation}
 $V^{\rm M}$ is accounting for thermal corrections to the Majorana mass. The third term, $V^{\rm D}$, in \eqref{eq:HamT} is
 the already accounted for effect of Dirac Yukawa interactions in \eqref{eq:V}-\eqref{eq:GammaDG}. 
 
 In summary, the new effects on Majorana masses are taken into account automatically with making the substitution in the source term
 \eqref{eq:DGsource}:
 \begin{equation}
 \label{eq:subs1}
\frac{M_{ii}^2-M_{jj}^2}{2 T} \,\longrightarrow\, \frac{1}{2T}\left(
(|Y^{\rm M}|^2_{ii} - |Y^{\rm M}|^2_{jj}) ( |\langle \phi \rangle|^2\,\, \Theta(\langle |\phi |\rangle-T)\,+\, 
2T(V^{\rm M}_{ii} -V^{\rm M}_{jj}
\right)\,,
 \end{equation}
 which amounts to 
 \begin{equation}
 \label{eq:subs}
\Delta M^2 := \Delta M_0^2\,\,\longrightarrow\,\, \Delta M^2(T) :=
\Delta |Y^{\rm M}|^2\left( |\langle \phi \rangle|^2\,\, \Theta(\langle |\phi |\rangle-T)\,+\, 
\frac{1}{16}\,T^2
\right)\,,
 \end{equation}
 where the zero-temperature contribution is $\Delta M_0^2$ which can also be written as $\Delta |Y^{\rm M}|^2 |\langle \phi \rangle|^2.$
We further note that the $Z'$ contributions to $V^{\rm M}$ are flavour-independent
and cancel out in $\Delta M^2(T)$. 

\begin{figure}[t!]
\begin{center}
\begin{tabular}{cc}
\hspace{-.4cm}
\includegraphics[width=0.5\textwidth]{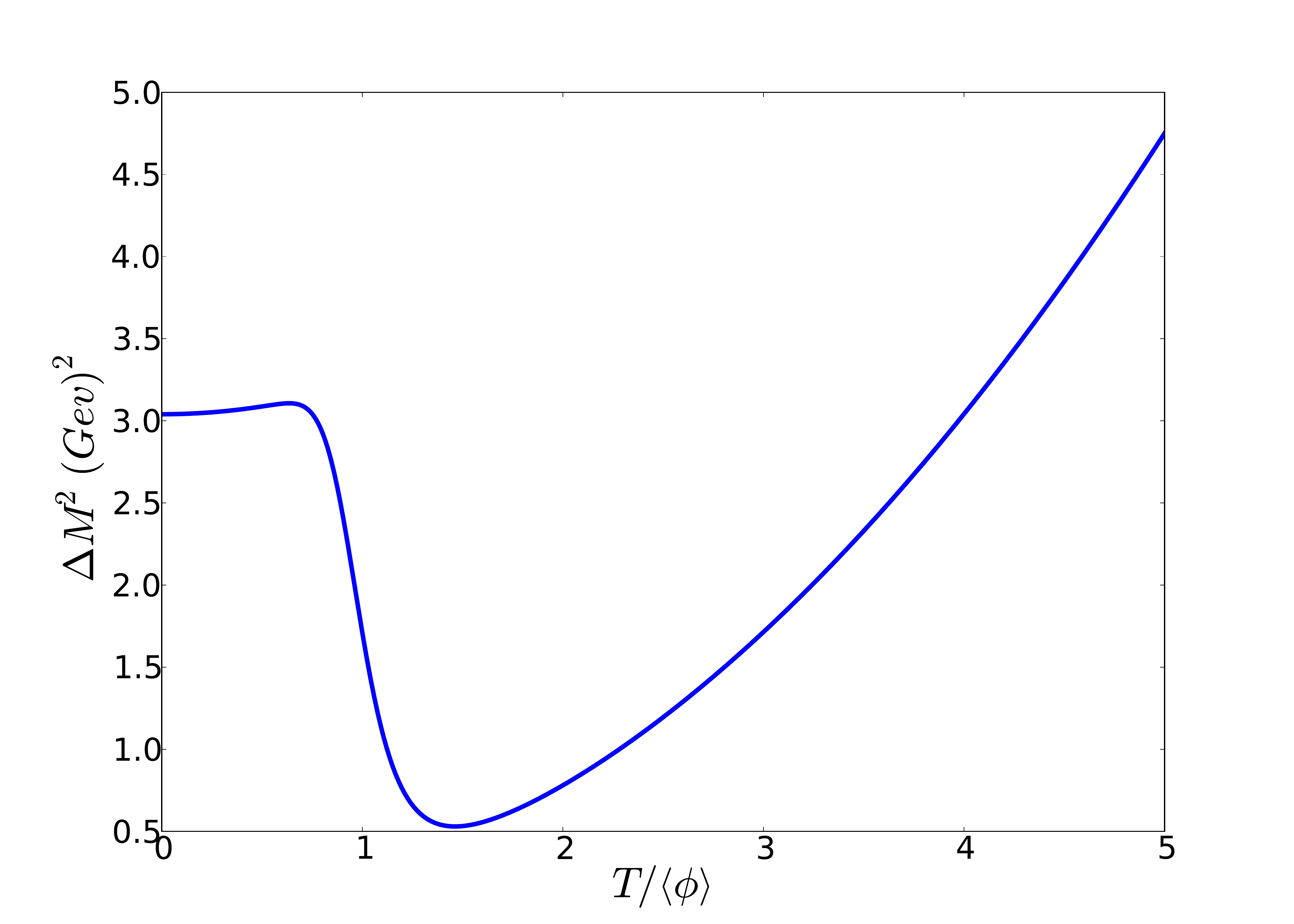}
&
\includegraphics[width=0.5\textwidth]{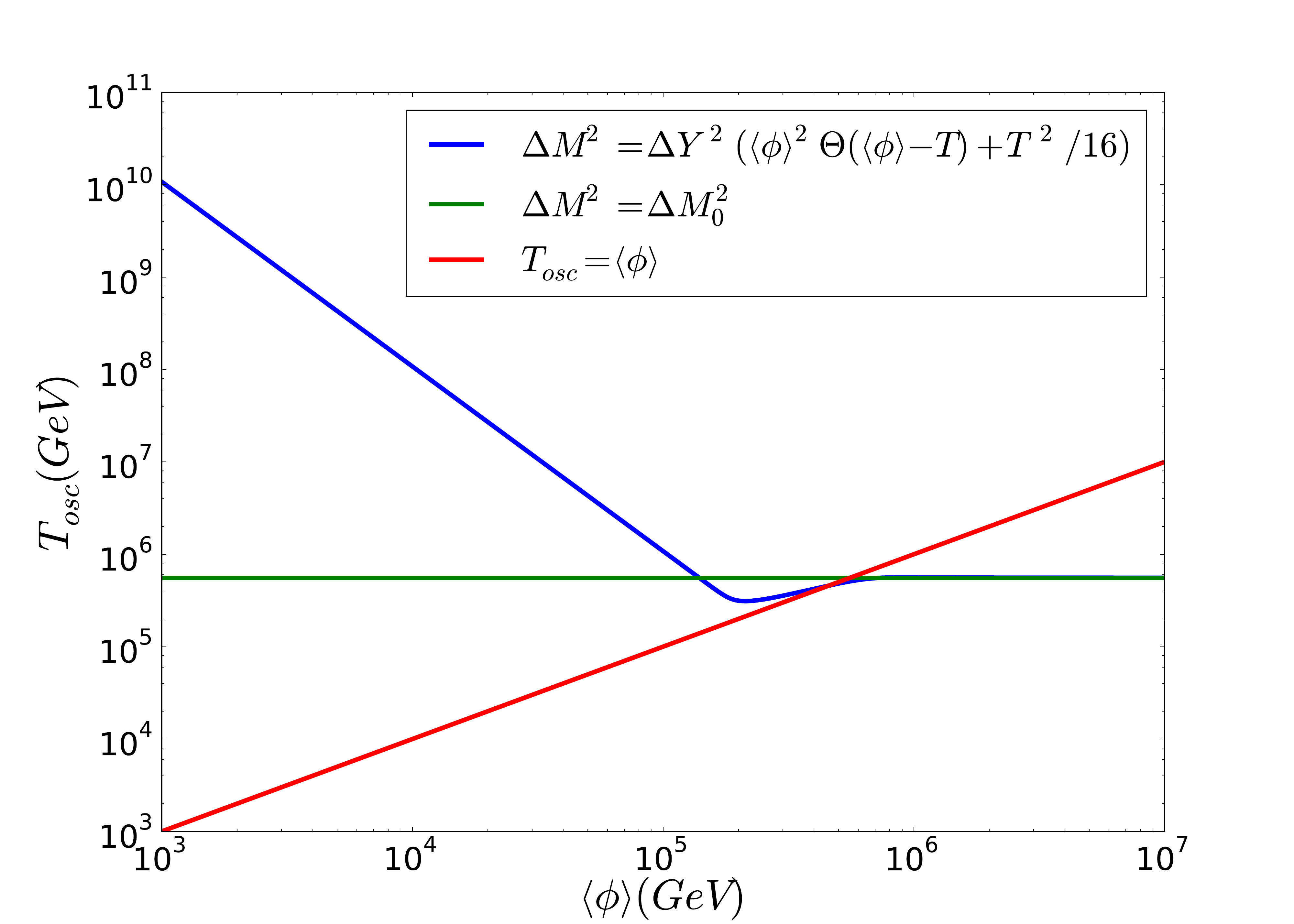}
\\
\ref{fig:0}(a) & \ref{fig:0}(b)  
\end{tabular}
\end{center}
\vskip-.4cm
\caption{
\label{fig:0}
\it \small Left panel shows the effective thermal mass squared difference $\Delta M^2(T)$ given by  \eqref{eq:subs} 
with smoothened theta function 
(and the initial value taken to be $\Delta M^2_0=3\, {\rm GeV}^2$)
as the function of the temperature over $\langle |\phi |\rangle$.  On the right panel, the blue curve
sketches the initial temperature $T_{\rm osc}$ as the function of $\langle |\phi |\rangle$ showing the transition between the 
unbroken ($T_{\rm osc}>\langle |\phi |\rangle$) and the broken ($T_{\rm osc}<\langle |\phi |\rangle$) phase.
The horizontal green line gives the value of $T_{\rm osc}$ computed in the regime 
of \cite{Drewes:2012ma} via \eqref{eq:Tosc}.
On the right of the plot, the blue and green curves coincide.  }
\end{figure}

On Fig.~\ref{fig:0}(a) we plot the effective $\Delta M^2(T)$ given by the right hand
side of \eqref{eq:subs} as the function of temperature. For future convenience we have smoothened the step-function
to account for a more physical behaviour. Essentially, the non-vanishing mass in the broken phase on the left is connected
at $T/\langle |\phi |\rangle \sim 1$
by a finite-width bubble wall to the unbroken phase where the mass receives only the purely thermal contribution.

The integral of the source term \eqref{eq:DGsource} 
in the original DG formulation ($z=T_{EW}/T$),
 \begin{equation}
  \label{eq:DGint} 
 {\rm DG:} \qquad
 \frac{M_{\rm Pl} T_{EW}}{\Delta M^2}\left(\int_{z_{\rm osc}}^1\frac{dz}{z^2}\right)\, ,
 \end{equation}
after the
substitution \eqref{eq:subs} becomes -- assuming $T_{\rm osc} > T_{B-L} = \langle |\phi |\rangle$,
 \begin{equation}
  \label{eq:OURint} 
 \frac{M_{\rm Pl} T_{EW}}{\Delta |Y^{\rm M}|^2}\left(
 \int_{z_{\rm osc}}^{T_{EW} /\langle |\phi |\rangle} \frac{dz}{T_{EW}^2/16}
 \,+
  \int_{T_{EW} /\langle |\phi |\rangle}^1 \frac{dz}{|\langle \phi \rangle |^2 z^2 \,+\, T_{EW}^2/16}
\right)
\end{equation}
or 
\begin{equation}
  \label{eq:OURint2} 
 \frac{M_{\rm Pl} T_{EW}}{\Delta |Y^{\rm M}|^2}\left(
  \int_{z_{\rm osc}}^1 \frac{dz}{|\langle \phi \rangle |^2 z^2 \,+\, T_{EW}^2/16}
\right)
\end{equation}
in the opposite case.

The initial `time' $z_{\rm osc}$ is determined in the similar manner to what was done before in \eqref{eq:Tosc}. For the 
$T_{\rm osc} > \langle |\phi |\rangle$ case it follows from
\begin{equation}
  \label{eq:Tosc2}
 \frac{1}{16}     \frac{\Delta |Y^{\rm M}|^2 \,  T_{\rm osc}^2}{2 T_{\rm osc}} \,=\, 2\pi \, H(T_{\rm osc}) \quad => \quad
 z_{\rm osc}\,=\, 64\pi   \frac{T_{EW}}{\Delta |Y^{\rm M}|^2\,M_{\rm Pl}^*}\,.
\end{equation}

Our result for the lepton flavour asymmetry is
\begin{equation}
  \label{eq:deltaLOUR}
\frac{n_{La}}{s} \,=\, - \gamma_{av} ^2 \times 7.3 \times 10^{-4} \sum_{c}\sum_{i\neq j} i \, 
 (Y^{\rm D \,\dagger}_{ai} Y^{\rm D}_{ic} Y^{\rm D \,\dagger}_{cj} Y^{\rm D }_{ja} -
 Y^{\rm D \,t}_{ai} Y^{\rm D\, *}_{ic} Y^{\rm D \,t}_{cj} Y^{\rm D \, *}_{ja})\, \times {\cal I}_{ij}\, ,
 \end{equation}
where ${\cal I}_{ij}$ is the integral in \eqref{eq:OURint},
\begin{equation}
  \label{eq:Iij} 
 {\cal I}_{ij}\,=\, 
  \frac{16}{ \sum_k (Y^{\rm M \,\dagger}_{ik} Y^{\rm M}_{ki} -Y^{\rm M \,\dagger}_{jk} Y^{\rm M}_{kj})}\,
  \frac{M_{\rm Pl} }{\langle |\phi |\rangle}\,
  \left( 1 -  \frac{\langle |\phi |\rangle}{T_{\rm osc}} +\frac{1}{4} {\rm tan}^{-1}\left(\frac{4 \langle |\phi |\rangle}{T_{EW}}\right) -
 \frac{1}{4} {\rm tan}^{-1}\left(4\right) 
  \right)\,,
 \end{equation}
 where
 \begin{equation}
 \label{eq:3.30}
 \langle |\phi |\rangle\, <\, T_{\rm osc} := \frac{\Delta |Y^{\rm M}|^2\,M_{\rm Pl}^*}{64 \pi}\,=\, 
 \frac{\Delta |M_0|^2\,M_{\rm Pl}^*}{64 \pi\,\langle |\phi |\rangle^2 }
 \,.
 \end{equation}
 
 \bigskip
 
The low-temperature case \eqref{eq:OURint} is treated similarly.
We note that in the case where $T_{\rm osc} $ approaches  $\langle |\phi |\rangle$ (or falls below it),  the first integral in \eqref{eq:OURint} 
disappears, since $T_{EW} / \langle |\phi |\rangle \to z_{\rm osc} $, in agreement with \eqref{eq:OURint2}. This is manifested
by the cancellation between the first and the second term inside the brackets in \eqref{eq:Iij}, so that
\begin{equation}
  \label{eq:Iij2} 
T_{\rm osc} \le \langle |\phi |\rangle: \quad
 {\cal I}_{ij}\,=\, 
  \frac{4}{ \sum_k (Y^{\rm M \,\dagger}_{ik} Y^{\rm M}_{ki} -Y^{\rm M \,\dagger}_{jk} Y^{\rm M}_{kj})}\,
  \frac{M_{\rm Pl} }{\langle |\phi |\rangle}\,
  \left(  {\rm tan}^{-1}\left(\frac{4 \langle |\phi |\rangle }{T_{EW}}\right) -
{\rm tan}^{-1}\left(
\frac{4 \langle |\phi |\rangle }{T_{\rm osc}}
\right) 
  \right)\, ,
 \end{equation}
 with $T_{\rm osc}$ in this case given by 
\eqref{eq:Tosc}.

The dependence of $T_{\rm osc}$ on the value of $\langle |\phi |\rangle$ is plotted on the right panel of
Fig.~\ref{fig:0} in blue. The red diagonal line is the $T_{\rm osc}=\langle |\phi |\rangle$ boundary separating
the broken from the unbroken phase. The horizontal green line gives the value of $T_{\rm osc}$ in the regime 
of \cite{Drewes:2012ma} given by \eqref{eq:Tosc}. It is valid for low temperatures (high vevs) 
$T_{\rm osc} \le \langle |\phi |\rangle$ i.e. to the right of the diagonal red line where the blue line coincides with the horizontal green line.
On the other hand, at high temperatures, $T_{\rm osc}>\langle |\phi |\rangle$, the blue line depicting $T_{\rm osc}$ 
is determined by the right hand side of Eq.~\eqref{eq:3.30}. In the transitional region where all three lines meet, the blue line of
$T_{\rm osc}$ briefly drops below the green line prediction of \cite{Drewes:2012ma}. This dip is a consequence of the local minimum 
on Fig.~\ref{fig:0}(a) which corresponds to the drop in the effective mass squared when one passes from the broken to the unbroken phase.

\section{Baryon asymmetry and phenomenology}
\label{sec:4}

Equations \eqref{eq:deltaLOUR},\eqref{eq:Iij} derived in the previous Section compute the lepton 
flavour asymmetry generated in the classically conformal 
 Standard Model $\times$ CW$_{B-L}$.  Electroweak sphalerons 
 process this lepton flavour asymmetry into baryon asymmetry of the Universe (BAU). 
 As explained in Sec. {\bf \ref{sec:3.1}}  in order to achieve a
 non-vanishing value of BAU it is required that at the time of electroweak phase transition, two of the
 flavours of sterile neutrinos are equilibrated with their SM decay products, but
 it is essential that the remaining flavour is not. Thus if the inequalities \eqref{eq:ARSequil} are satisfied,
 the BAU is produced $\sim - \,n_{Le}$. The estimate \cite{Drewes:2012ma} is 
 \begin{equation}
  \label{eq:BAU} 
\frac{n_{b}}{s} \,\simeq\, -\frac{3}{14} \times 0.35 \times \frac{n_{Le}}{s}
\end{equation}
where the observed value of the asymmetry is $n_b^{\rm obs}/s = (8.75 \pm 0.23)\times 10^{-11}$.

In what follows we would like to determine the range of our model parameters for which the required baryon asymmetry is achieved.
In the neutrino sector we use the standard Casas-Ibarra parametrisation \cite{Casas:2001sr} of the see-saw Dirac Yukawa couplings, 
\begin{equation}
  \label{eq:mixing} 
Y^{D\, \dagger} \,=\, U_{\nu} \cdot \sqrt{m_{\nu}} \cdot {\cal R} \cdot \sqrt{M} \times \frac{\sqrt{2}}{v}\,,
\end{equation}
where $m_\nu$ and $M$ are diagonal masses of active and Majorana neutrinos respectively, and $v=246$ GeV.
The active-neutrino-mixing matrix $U_{\nu}$ is the PMNS matrix which contains six real parameters, including three measured mixing angles 
and three CP-phases. The matrix ${\cal R}$ is parametrized by three complex angles $\omega_{ij}$.  

In our analysis we will choose and fix the values of $m_\nu$ consistent with the solar and atmospheric neutrino mass differences.
We will further fix a generic value for the three CP-phases. The unknown complex angles $\omega_{ij}$ should be varied over the parameter space. To keep things as simple as possible, we will chose a 2-dimensional subspace on which we vary the real and imaginary
 parts of the complex angle $\omega_{23}$, while keeping $\omega_{12}$ and $\omega_{13}$ fixed. 
 For easy comparison, our 2-d slice of the $omega$-space is the same as in Ref.~\cite{Drewes:2012ma} (it can be read off
 Scenarios 1-3 and 5-7 in Tables~\ref{tab:1}  and \ref{tab:2} below). 
 For completeness, we will also comment on the results of varying the other complex angles and CP phases of the parameter space.
 
 Different choices of the
 three  Majorana masses will characterise different benchmark points we consider. Since in our case the Majorana particle mass (we drop the subscript $0$ in
 what follows) is 
 $M= Y^{\rm M} \langle \phi \rangle$, there is an additional scale $\langle \phi \rangle$ which we will vary and specify\footnote{Phenomenologically,
 it makes sense to use  $\langle \phi \rangle$ and $M$ as the two independent parameters,
  rather than, say $Y^{\rm M}$ and $ \langle \phi \rangle$.}. 

\begin{figure}[t!]
\begin{center}
\begin{tabular}{cc}
\hspace{-.4cm}
\includegraphics[width=0.5\textwidth]{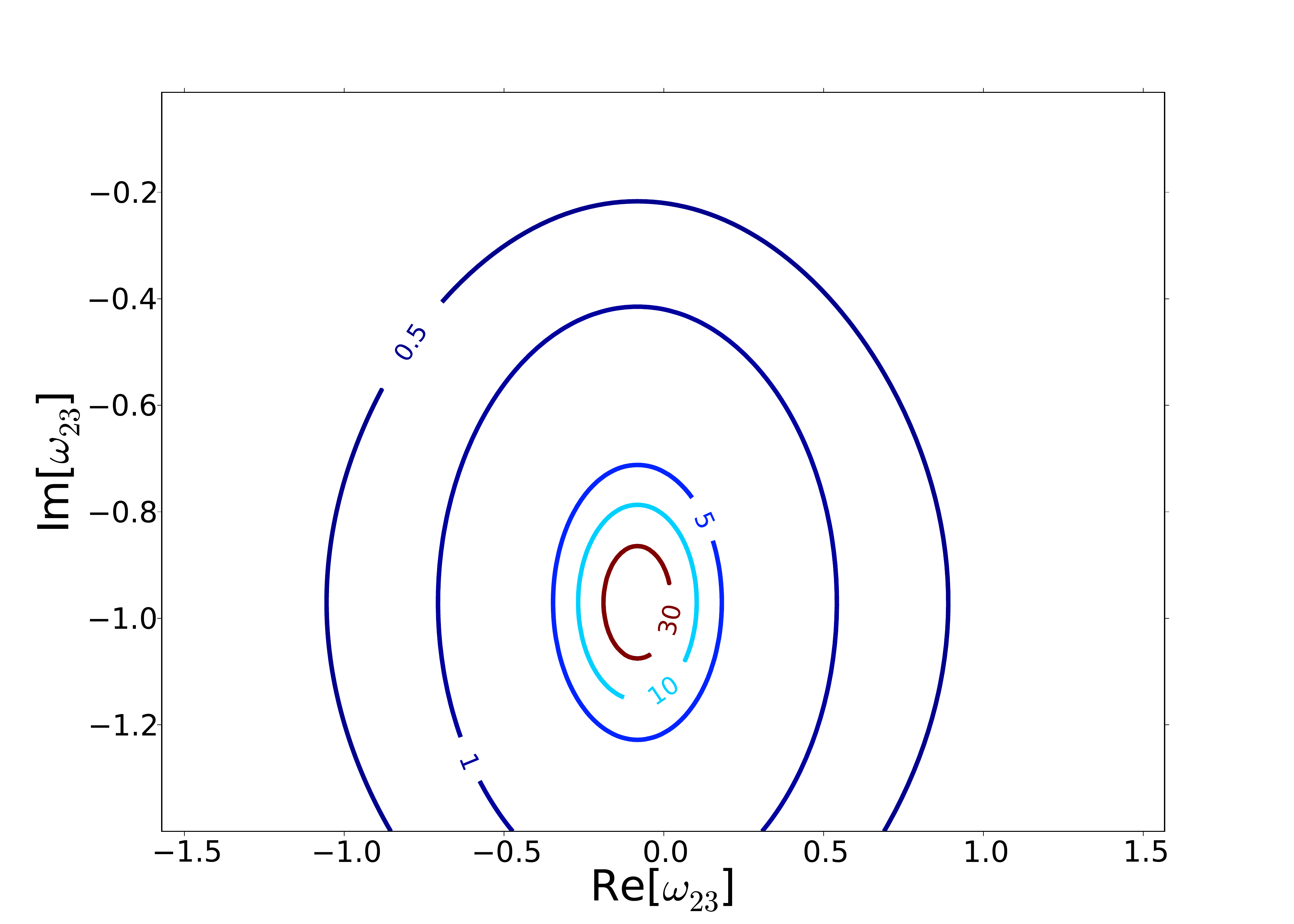}
&
\includegraphics[width=0.5\textwidth]{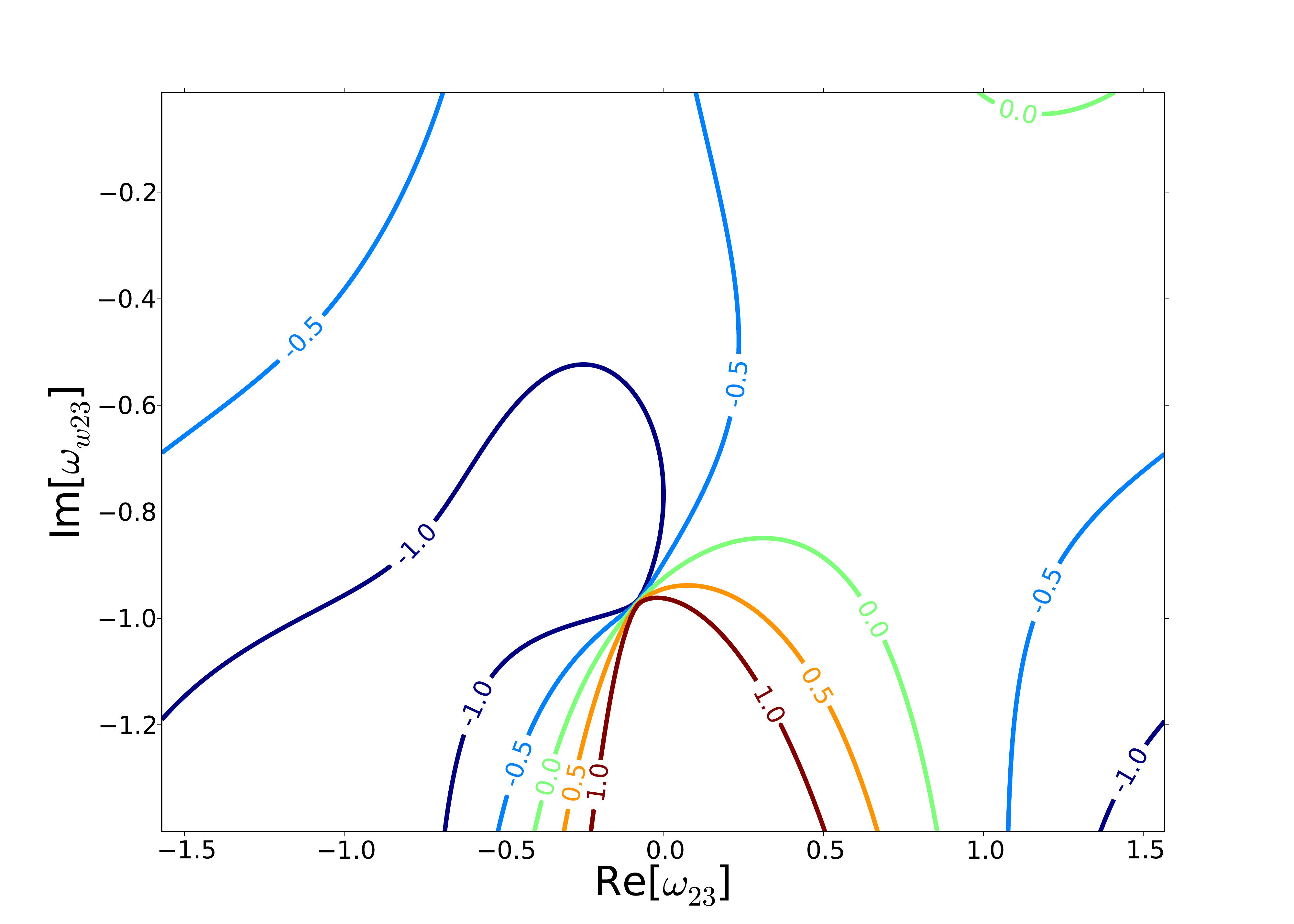}
\\
\ref{fig:1}(a) $M$(GeV) & \ref{fig:1}(b)  ${n_{b}/}{n_{b}^{\rm obs}}$
\end{tabular}
\end{center}
\vskip-.4cm
\caption{
\label{fig:1}
\it \small Left panel shows maximal values of Majorana masses in GeV for which the wash-out bound in Eq.~\eqref{eq:washout}
can be achieved. The panel on the right shows contours for the baryon asymmetry produced, normalised to the observed value.
Majorana masses used in \ref{fig:1}(b) are taken from \ref{fig:1}(a) for each value of $Re[\omega_{23}]$ and $Im[\omega_{23}]$.
In both plots we vary $Re[\omega_{23}]$ and $Im[\omega_{23}]$ keeping other parameters of the model fixed at indicative values 
as in Ref.~\cite{Drewes:2012ma}, detailed in the Tables~\ref{tab:1}  and \ref{tab:2}.
}
\end{figure}
\begin{figure}[h!]
\begin{center}

\hspace{-.4cm}
\includegraphics[width=0.6\textwidth]{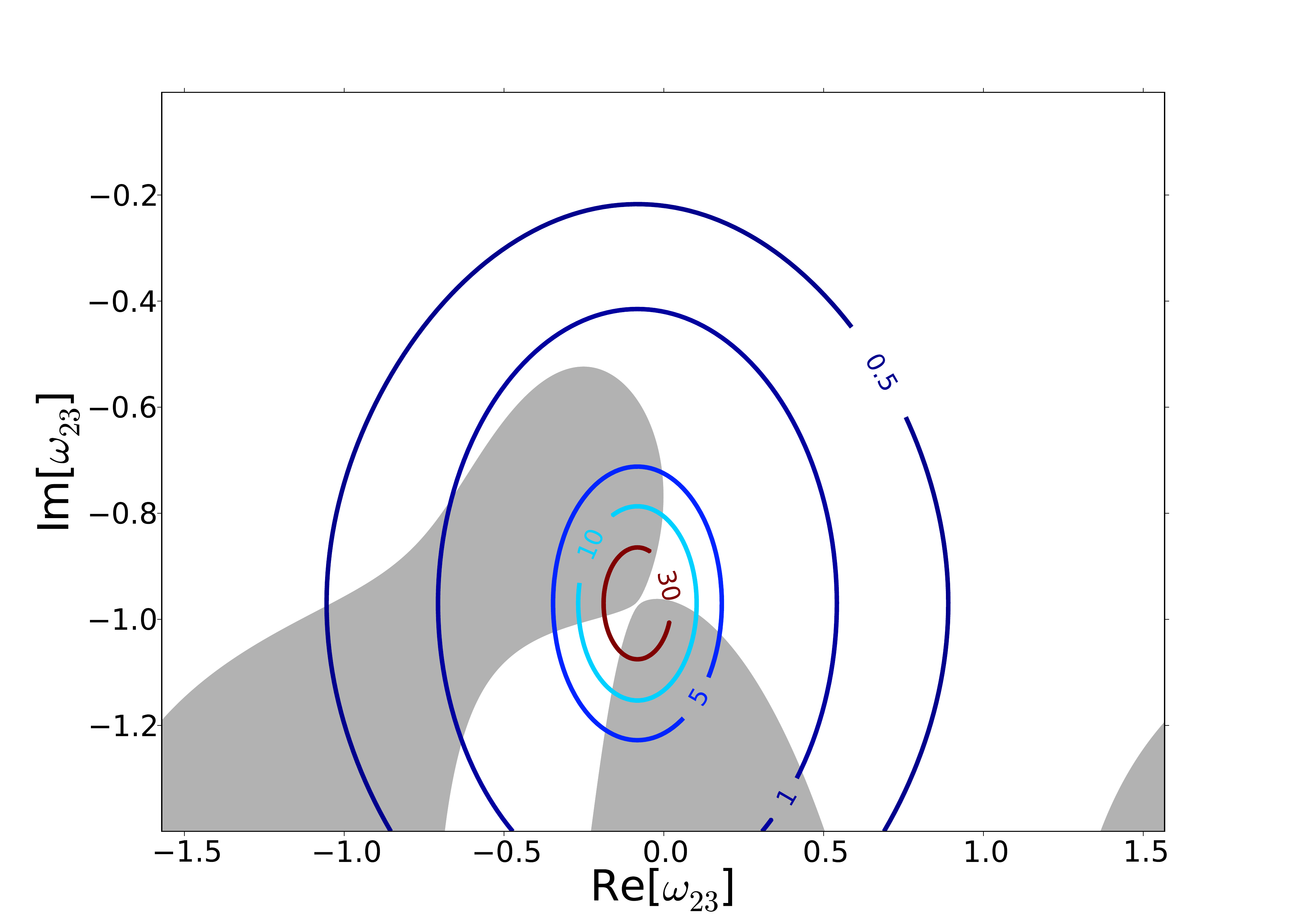}

\end{center}
\vskip-.4cm
\caption{
  \label{fig:1sup}
\it \small Superposition of the Majorana mass contours in GeV 
satisfying the wash-out bound with the baryon asymmetry produced with shaded regions denoting
the required baryon asymmetry from Fig.~\ref{fig:1}
}
\end{figure}

First we would like to determine the range of allowed values of Majorana masses for sterile neutrinos in our model.
The wash-out rates for the lepton flavours $a=e,\mu,\tau$ are given by $\Gamma_a$, and we require that 
\begin{equation}
  \label{eq:washout}
\frac{\Gamma_e}{H(T_{EW})}\,:=\, \frac{1}{2}\sum_i
Y^{\rm D \,\dagger}_{ei}  Y^{\rm D }_{ie} \,\gamma_{av} \, \frac{T_{EW}}{H(T_{EW})}\, <\, 1 \,.
 \end{equation}
Figure \ref{fig:1}(a) shows mass contours in GeV of the lightest Majorana neutrino flavour, such that the
wash-out rate =1 is achieved. This can be interpreted as an upper bound on Majorana masses for which \eqref{eq:washout}
is satisfied. Quite clearly from this perspective it is straightforward to realise $M$ in the region from 
few 100 MeV to above 30 GeV or even up to a TeV. The BBN constrains the lower limit to $M> 200$ MeV, so we have
\begin{equation}
  \label{eq:range}
200 \,{\rm MeV}\, <\, M \,\lesssim\, {\rm few}\times 100 \,{\rm GeV} \,.
 \end{equation}

What about the produced baryon asymmetry? Figure \ref{fig:1}(b) plots the ratio ${n_{b}/}{n_{b}^{\rm obs}}$.
The baryon asymmetry here is computed using Eq.~\eqref{eq:deltaLDG} 
derived for the simple Majorana mass model \cite{Drewes:2012ma}, where the values of $M$ at each point on the parameter space are 
taken from Fig.~\ref{fig:1}(b).
Below we will also compute the asymmetry in the 
classically conformal Standard Model $\times$ CW$_{B-L}$.
To generate the observed asymmetry we need to be inside the +1 or -1 contours in Fig.~\ref{fig:1}(b). 

Figure~\ref{fig:1sup} depicts the superposition of the two panels
of Fig.~\ref{fig:1}. It can be seen that the required baryon asymmetry (the area inside the two shaded contours in 
 Fig.~\ref{fig:1sup}) is indeed generated in the above mass range.
 
 \medskip
 Figures~\ref{fig:1}, \ref{fig:1sup} were obtained by varying the real and imaginary parts of $\omega_{23}$ while keeping
 other parameters fixed. We have also checked that desired amounts of the wash-out and the baryon asymmetry are produced in
 sizable regions of the parameter space when other complex angles and CP phases are varied. In our benchmark points 
 described in the Tables below, the fixed parameters were chosen inside these regions.
 
 \medskip

\begin{table}[th!]
\begin{tabular}{|l| c| c| c|| c| }
\hline
& Scenario 1 & Scenario 2 & Scenario 3 & Scenario 4 \\ \hline
$M_1$& $0.5\,{\rm GeV}$ & $3.6\,{\rm GeV}$ & $200.0\,{\rm GeV}$ & $1.0\,{\rm GeV}$ \\
$M_2$& $0.6\,{\rm GeV}$ & $4.0\,{\rm GeV}$ & $250.0\,{\rm GeV}$ & $2.0\,{\rm GeV}$ \\
$M_3$& $0.7\,{\rm GeV}$ & $4.4\,{\rm GeV}$ & $300.0\,{\rm GeV}$ & $3.0\,{\rm GeV}$ \\
\hline
$m_1$& $0.0\,{\rm meV}$ & $0.0\,{\rm meV}$ & $0.0\,{\rm meV}$ & $2.5\,{\rm meV}$ \\
$m_2$& $8.7\,{\rm meV}$ & $8.7\,{\rm meV}$ & $8.7\,{\rm meV}$ & $9.1\,{\rm meV}$ \\
$m_3$& $49.0\,{\rm meV}$ & $49.0\,{\rm meV}$ & $49.0\,{\rm meV}$ & $49.0\,{\rm meV}$ \\
$s_{12}$& 0.55 & 0.55 & 0.55 & 0.55 \\
$s_{23}$& 0.63 & 0.63 & 0.63 & 0.63 \\
$s_{13}$& 0.16 & 0.16 & 0.16 & 0.16 \\
$\delta$& $-\pi/4$ &$ -\pi/4$ & $-\pi/4$ &$ \pi$ \\
$\alpha_1$& 0 & 0 & 0 & $-\pi$ \\
$\alpha_2$& $-\pi/2$ &$ -\pi/2$&$ -\pi/2$ &$ \pi$ \\
$\omega_{12}$& 1+2.6i & 1+2.6i & 1+2.6i & -1+1.5i \\
$\omega_{13}$& 0.9+2.7i & 0.9+2.7i & 0.9+2.7i & 0.5+2.6i \\
$\omega_{23}$& 0.3-1.5i & -1.2i & -0.05-0.975i &$ \pi$-2.4i \\
\hline
$n_{Le}/(s\times 2.5 \times 10^{-10})$& -4.4 & -6.7 & -5 & -8.3 \\
$n_{L\mu}/(s\times 2.5 \times 10^{-10})$& 39 & 32 & 108 & 32 \\
$n_{L\tau}/(s\times 2.5 \times 10^{-10})$& -34 & -25 & -103 & -24 \\
$\Gamma_e/H(T_{EW})$& 0.68 & 0.64 & 0.84 & 0.59 \\
$\Gamma_\mu/H(T_{EW})$& 68 & 290 & $1\times 10^4$ & 410 \\
$\Gamma_\tau/H(T_{EW})$& 220 & 920 &$ 4\times 10^4$ & 150 \\
$T_{osc}$& $2\times 10^5 \,{\rm GeV}$&$ 5\times10^5\,{\rm GeV}$ &$ 10^7\,{\rm GeV}$ & $5\times 10^5\,{\rm GeV}$ \\
\hline
\end{tabular}
\caption{
  \label{tab:1}
\it \small Four benchmark points corresponding to different ranges of Majorana masses.}

\end{table}

\begin{figure}[t!]
\begin{center}
\begin{tabular}{cc}
\hspace{-.4cm}
\includegraphics[width=0.5\textwidth]{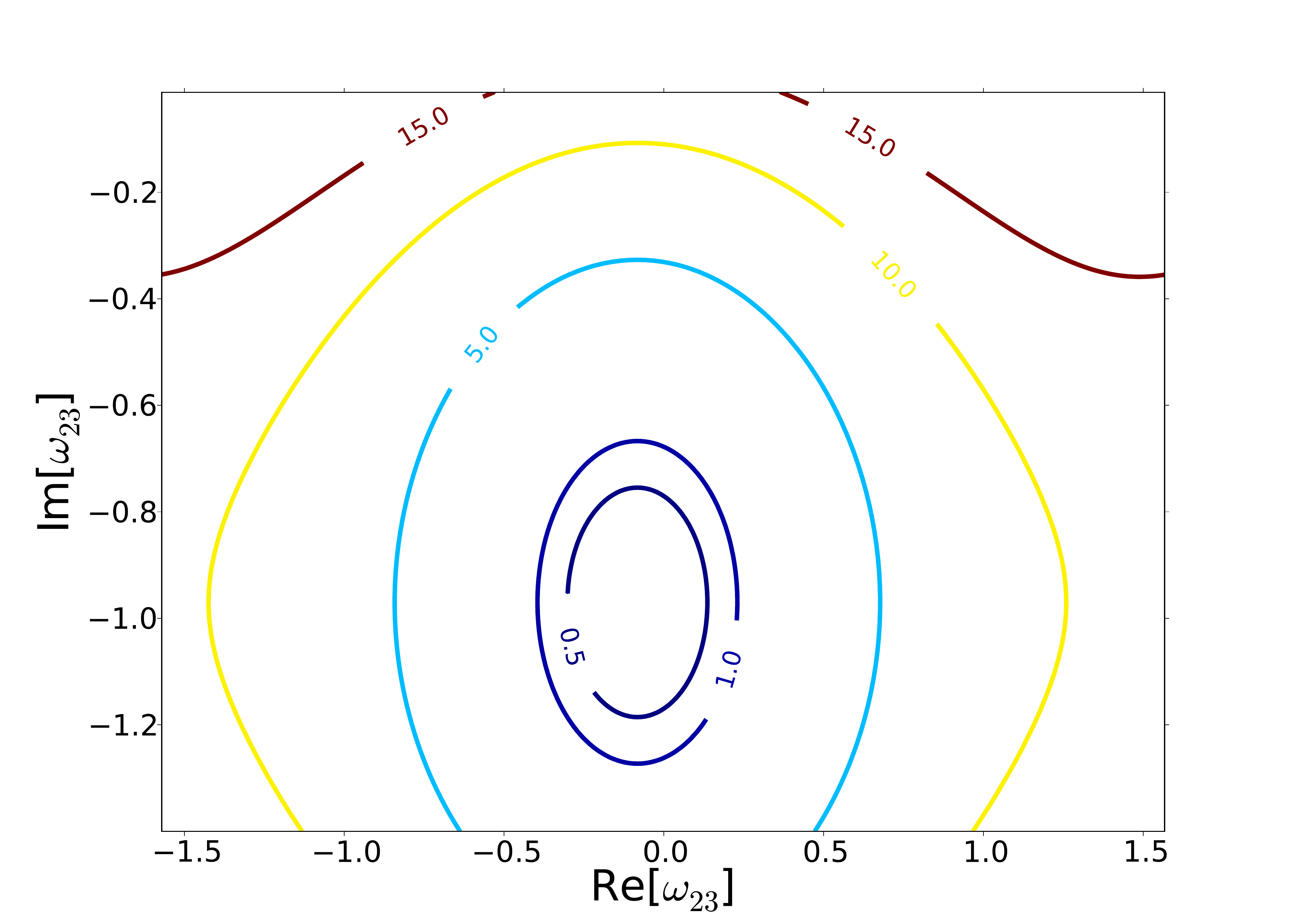}
&
\includegraphics[width=0.5\textwidth]{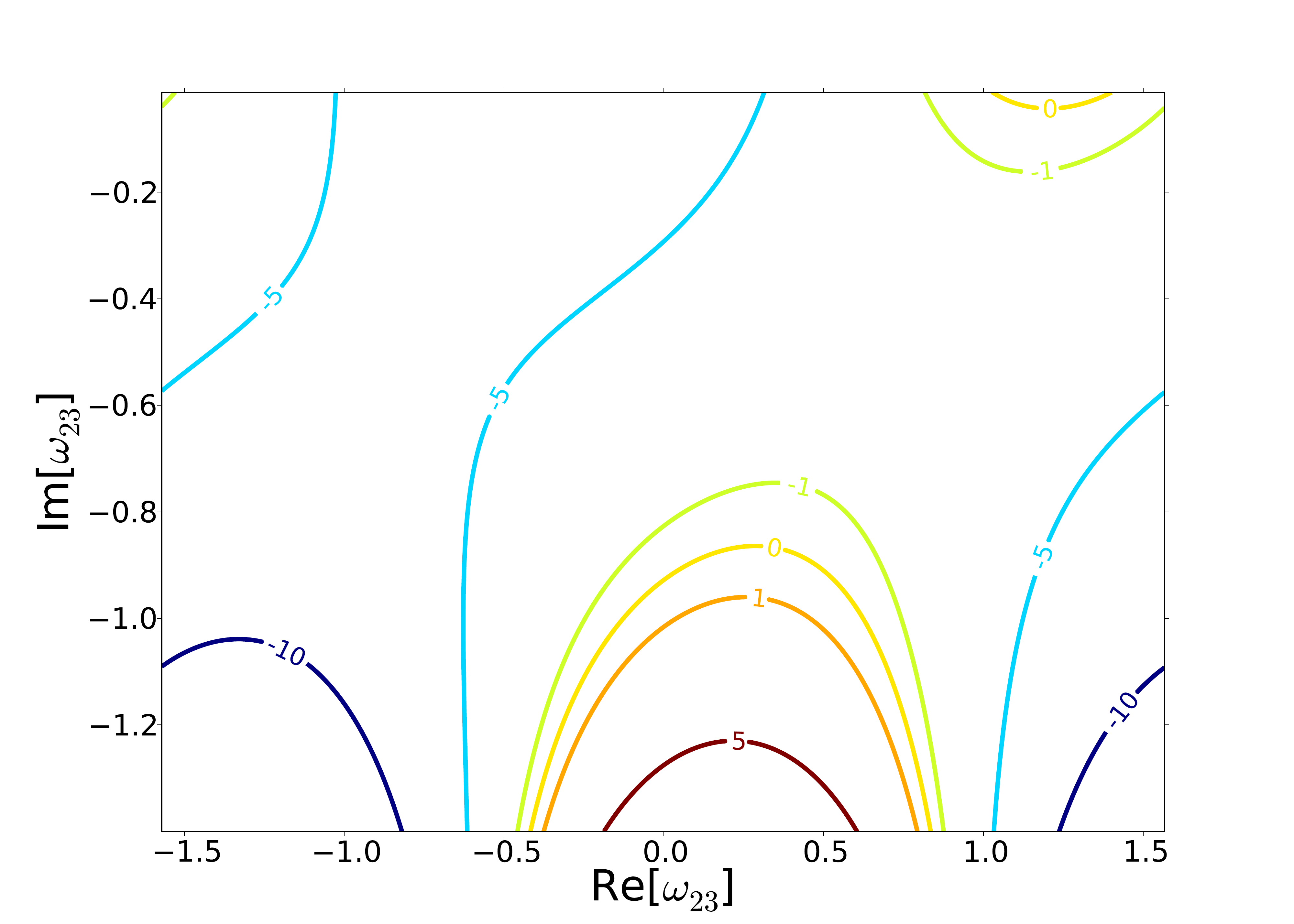}
\\
\ref{fig:2}(a) Wash-out & \ref{fig:2}(b)  ${n_{b}/}{n_{b}^{\rm obs}}$
\end{tabular}
\end{center}
\vskip-.4cm
\caption{
\label{fig:2}
\it \small The wash-out rate (left panel) and the normalised baryon asymmetry computed in the 
classically conformal $B-L$ model. The values of model parameters are defined in the text.
}
\end{figure}
\begin{figure}[h!]
\begin{center}

\hspace{-.4cm}
\includegraphics[width=0.6\textwidth]{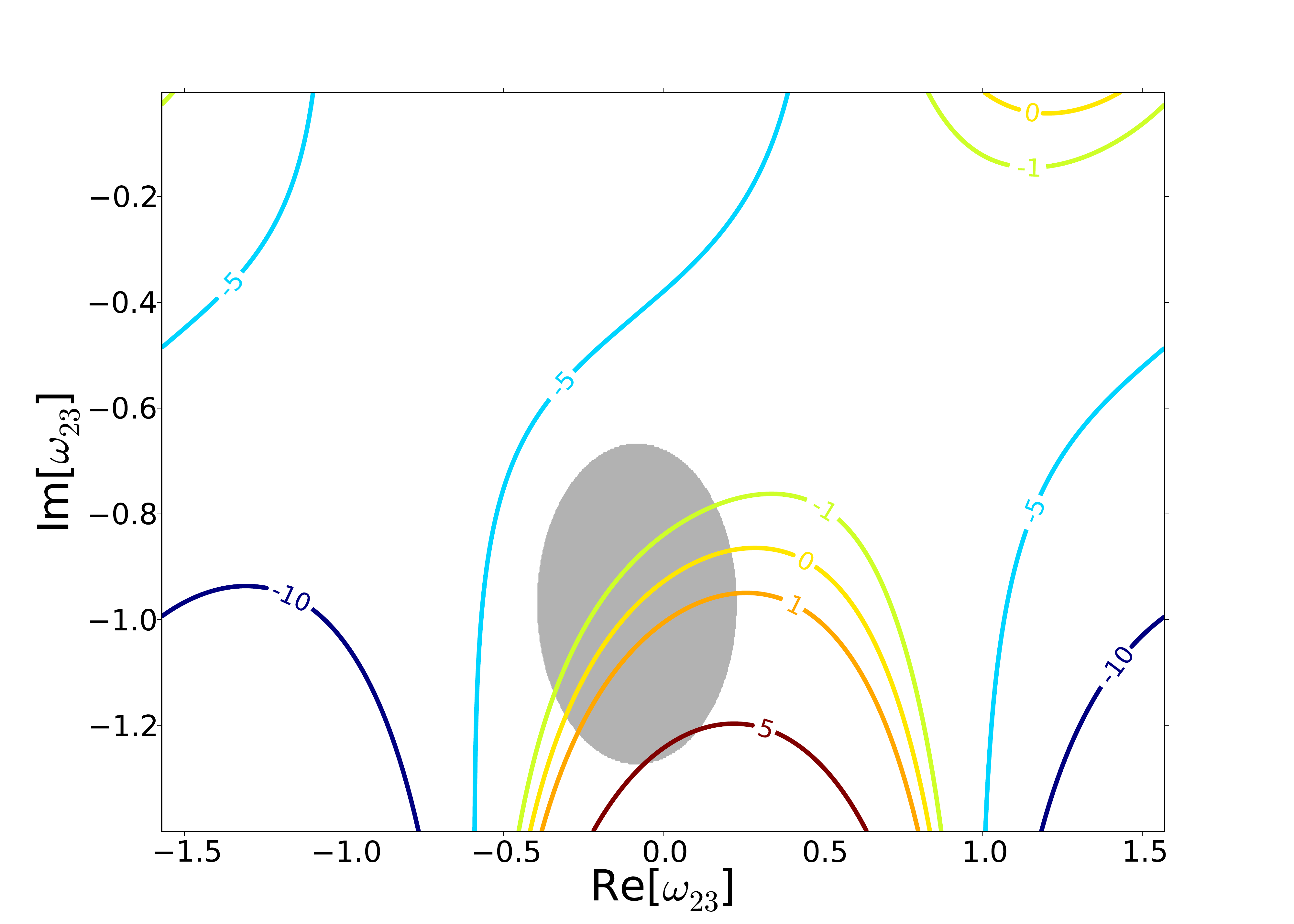}

\end{center}
\vskip-.4cm
\caption{
  \label{fig:2sup}
\it \small Superposition of the wash-out rate $\le 1$(inside the shaded ellipse) with the baryon asymmetry produced from Fig.~\ref{fig:2}
}
\end{figure}

In Table~\ref{tab:1} we present our first four benchmark point scenarios. The lepton flavour asymmetry 
$n_{La}/{s}$ with $a=e,\mu,\tau$ in all four cases in this Table
is calculated using Eq.~\eqref{eq:deltaLDG} in the simple Majorana mass model in the formalism of  \cite{Drewes:2012ma}.
We also show the washout rates for the three flavours, $\Gamma_a/H(T_{EW})$ and the value of the oscillation temperature.

In Scenarios 1, 2 $\&$ 3 we vary Majorana masses $M_i$ of sterile neutrinos from $\sim 500$ MeV (Scenario 1) through $\sim 4$ GeV
(Scenario 2) to $\sim 200-300$ GeV (Scenario 3). For convenience values of active neutrino masses in these three scenarios are chosen to be the same 
as in Scenario I in \cite{Drewes:2012ma}. Same applies to the choices of mixing angles. The main lesson of these benchmarks is to demonstrate 
the range of variation of Majorana masses in \eqref{eq:range}.

The fourth Scenario in Table~\ref{tab:1} is included for completeness, it reproduces Scenario II of \cite{Drewes:2012ma} and
has a different selection of active neutrino mass values from Scenarios 1-3.

\bigskip
Having established the likely range of Majorana masses, 
we now proceed to our main point - namely the analysis of the the classically conformal Standard Model $\times$ CW$_{B-L}$ where 
the matter-anti-matter asymmetry is computed using the formalism of Section~{\bf \ref{sec:3.2}}.
 
The right panel of Figure \ref{fig:2} shows the baryon asymmetry (normalised to its observed value) 
computed using  Eqs.~\eqref{eq:deltaLOUR}-\eqref{eq:Iij}.
The values of Majorana masses are chosen in the GeV range: $M_1=3.6$ GeV, $M_2=4.0$ GeV and $M_3=4.4$ GeV,
precisely as in our Scenario 2 in Table~\ref{tab:1}. 
The value of the Coleman-Weinberg vev is chosen $\langle |\phi |\rangle = 10^5$ GeV which corresponds to the Higgs portal coupling
$\lambda_{P}=\frac{1}{2}\left(\frac{125\, {\rm GeV}}{\langle |\phi |\rangle}\right)^2 \simeq 0.78 \times 10^{-6}$. To achieve the required
BAU we must be either below the +1 contour or above the -1 curve. This amounts to almost the entire area of Fig.~\ref{fig:2}(b)
being available. 

The left plane, Fig.~\ref{fig:2}(a), plots the wash-out rate contours for the same choice of parameters. 
Here we have to be inside the +1 ellipse for baryogenesis to succeed. The superposition of this wash-out $\le 1$ contour
with the baryon asymmetry calculated and depicted on Fig.~\ref{fig:2}(b) is shown on Fig.~\ref{fig:2sup}.

\bigskip
In the above example we chose a relatively large CW vev, $\langle |\phi |\rangle = 10^5$ GeV, not much below the value of
$T_{\rm osc} = 5 \times 10^5$ GeV computed for these GeV-scale values of $M_i$'s. As the result we ended up with a rather
small value of the Higgs portal coupling, $\lambda_{P} \sim 10^{-6}$.

\begin{figure}
\begin{center}

\hspace{-.4cm}
\includegraphics[width=0.75\textwidth]{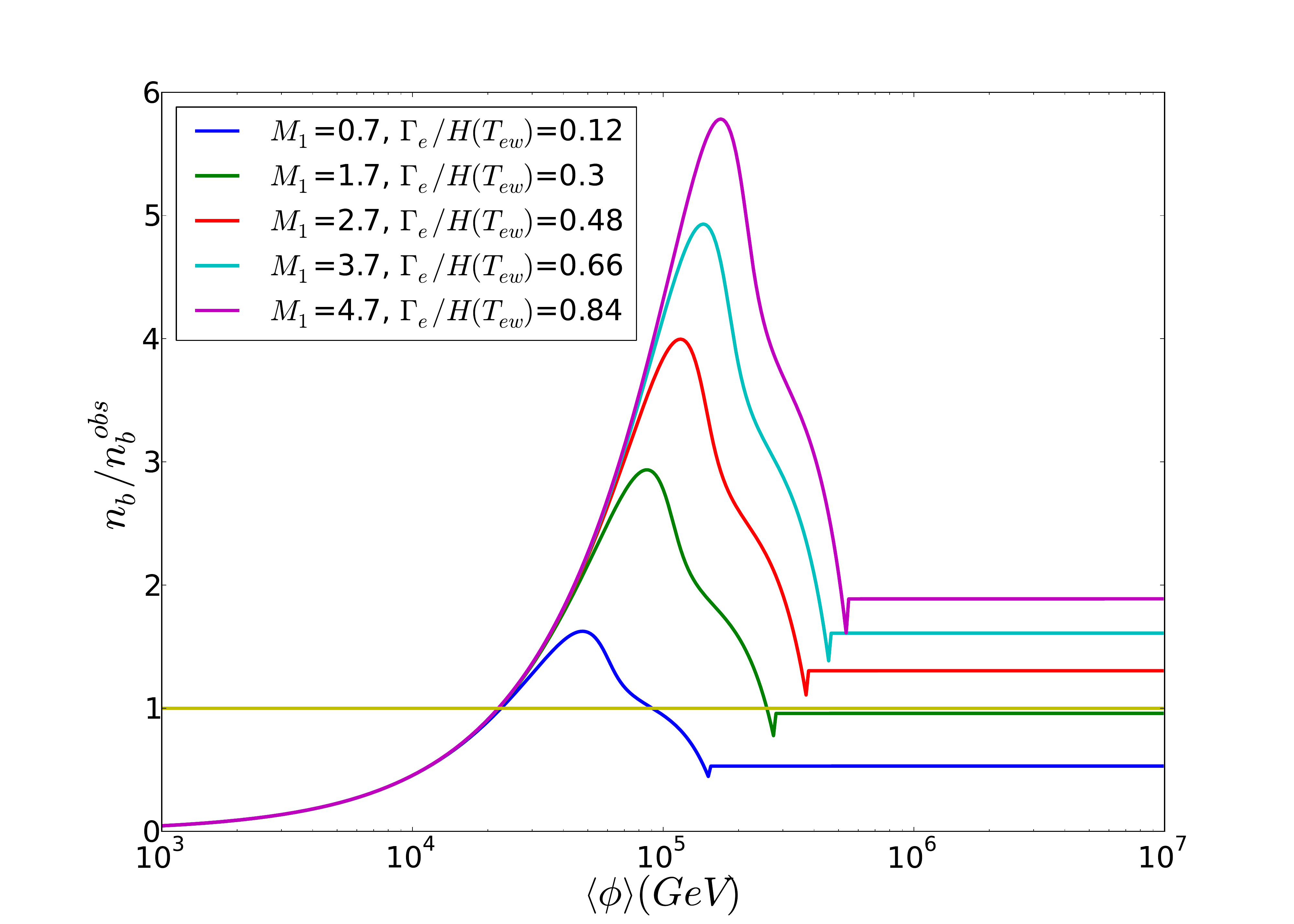}

\end{center}
\vskip-.4cm
\caption{
  \label{fig:3}
\it \small Baryon asymmetry (normalised by the observed value) as the
function of $\langle |\phi |\rangle$ for the range of masses between 0.7 GeV and 4.7 GeV. The washout rates for the electron neutrino
flavour (all less than 1 as required) are also shown in the legend.
}
\end{figure}

A natural and important question to ask is how much freedom 
do we have to lower $\langle |\phi |\rangle$ (and thus raise $\lambda_{P}$) while 
keeping other parameters, such as Majorana masses and consequentially $T_{\rm osc}$ fixed? 
Figure~\ref{fig:3} plots the baryon asymmetry (divided by the observed value) as the
function of $\langle |\phi |\rangle$ for the range of masses between 0.7 GeV and 4.7 GeV (from bottom to top). The figure also indicates 
the values of wash-out rates $\Gamma_e/H(T_{EW}) <1$.
The flat distributions on the right of the plot correspond to values of $\langle |\phi |\rangle$ reaching and exceeding the 
relevant values of temperature $T_{\rm osc}$ where leptogenesis begins. These constant values of the generated baryon asymmetry
agree with those computed using the non-dynamical Majorana masses in the formalism of \cite{Drewes:2012ma} reviewed 
above in Sec.~{\bf \ref{sec:3.1}}. 
To the left of the plateau on Fig.~\ref{fig:3} there is a small dip followed by a  broad peak which emerges largely 
due to the first integral in \eqref{eq:OURint}. The small dip is the reflection of the dip in the oscillation temperature 
$T_{\rm osc}$ in the middle of Fig.~\ref{fig:0}(b).

Finally, to the left of the plot on Fig.~\ref{fig:3}, at small values of $\langle |\phi |\rangle$, all contours converge and tend to 
zero rather uniformly, independently of the values of $M$'s.
To understand this point, note that according to  \eqref{eq:deltaLOUR},\eqref{eq:Iij},
\begin{equation}
  \label{eq:est} 
n_{b}\,\sim\, \frac{(Y^{\rm D})^4}{(Y^{\rm M})^2}\, \,\frac{M_{\rm Pl}}{\langle |\phi |\rangle}\, \sim\, 
\frac{\langle |\phi |\rangle\,m^2 M_{\rm Pl}}{v^4}\, \to \, 0\, \, , \quad {\rm as}\, \,\langle |\phi |\rangle\,\to \, 0 \,,
\end{equation}
and independently of $M$.

As the result, Fig.~\ref{fig:3} shows that rather independently of the values of the chosen Majorana masses,
the contours cross the observed value of baryon asymmetry (normalised at 1) for values of 
$\langle |\phi |\rangle \simeq 1.2 \times 10^{4}$ GeV. This gives $\lambda_{P} \simeq 0.5 \times 10^{-4}$.

\begin{table}[th!]
\begin{tabular}{|l| c| c| c| }
\hline
& Scenario 5 & Scenario 6 & Scenario 7 \\ \hline
$\left<\phi\right>$& $10^5\,{\rm GeV}$ & $2.5\times 10^4\,{\rm GeV}$ & $3.4\times 10^3\,{\rm GeV}$ \\
$M_1$& $3.6\,{\rm GeV}$ & $3.6\,{\rm GeV}$ & $3.96\,{\rm GeV}$ \\
$M_2$& $4.0\,{\rm GeV}$ & $4.0\,{\rm GeV}$ & $4.0\,{\rm GeV}$ \\
$M_3$& $4.4\,{\rm GeV}$ & $4.4\,{\rm GeV}$ & $4.04\,{\rm GeV}$ \\
\hline
$m_1$& $0.0\,{\rm meV}$ & $0.0\,{\rm meV}$ & $0.0\,{\rm meV}$ \\
$m_2$& $8.7\,{\rm meV}$ & $8.7\,{\rm meV}$ & $8.7\,{\rm meV}$ \\
$m_3$& $49.0\,{\rm meV}$ & $49.0\,{\rm meV}$ & $49.0\,{\rm meV}$ \\
$s_{12}$& 0.55 & 0.55 & 0.55 \\
$s_{23}$& 0.63 & 0.63 & 0.63 \\
$s_{13}$& 0.16 & 0.16 & 0.16 \\
$\delta$& $-\pi/4$ & $-\pi/4$ &$ -\pi/4$ \\
$\alpha_1$& 0 & 0 & 0 \\ 
$\alpha_2$&$-\pi/2$ & $-\pi/2$&$ -\pi/2$ \\
$\omega_{12}$& 1+2.6i & 1+2.6i & 1+2.6i  \\
$\omega_{13}$& 0.9+2.7i & 0.9+2.7i & 0.9+2.7i\\
$\omega_{23}$& 0.3-1.5i & -1.2i & -0.04-0.976i \\
\hline
$n_{Le}/(s\times 2.5 \times 10^{-10})$& -18 & -5 & -6.6 \\
$n_{L\mu}/(s\times 2.5 \times 10^{-10})$& 99 & 27 & 41 \\
$n_{L\tau}/(s\times 2.5 \times 10^{-10})$& -81 & -22 & -34 \\
$\Gamma_e/H(T_{EW})$& 0.64 & 0.64 & 0.67 \\
$\Gamma_\mu/H(T_{EW})$& 290 & 290 & 304 \\
$\Gamma_\tau/H(T_{EW})$& 920 & 920 & 960 \\
$T_{osc}$&$10^6\,{\rm GeV}$ & $7.5\times 10^7\,{\rm GeV}$ & $9.8\times 10^7\,{\rm GeV} $\\
\hline
\end{tabular}
\caption{
  \label{tab:2}
\it \small Three benchmark points in the classically conformal $B-L$ model corresponding to Majorana masses in the GeV range, 
with the values of the Coleman-Weinberg vev $\langle |\phi |\rangle\, =10^5, \,\, 2.5 \times 10^{4}$ and  $3.4 \times 10^3$ GeV.
}
\end{table}
\begin{table}[th!]
\begin{tabular}{|l| c| c| c| }
\hline
& Scenario 5 & Scenario 6 & Scenario 7 \\ \hline
$\left<\phi\right>$& $10^5\,{\rm GeV}$ & $2.5\times 10^4\,{\rm GeV}$ & $3.4 \times 10^3\,{\rm GeV}$ \\
\hline
$\lambda_p$& $8\times 10^{-7}$ &$10^{-5}$ &$ 0.7\times 10^{-3}$ \\
$Y^{\rm M}_1$& $3.6\times 10^{-5}$ &$ 1.4\times 10^{-4}$ & $1.2\times 10^{-3}$ \\
$Y^{\rm M}_2$&$ 4\times 10^{-5}$ &$ 1.6\times 10^{-4}$  &  $1.2\times 10^{-3}$ \\
$Y^{\rm M}_3$& $ 4\times 10^{-5}$ &$ 1.8\times 10^{-4}$ &  $1.2\times 10^{-3}$ \\
$\left<Y^{\rm D}\right>$& $4\times10^{-8}$ &$4\times10^{-8}$ & $4\times10^{-8}$ \\
$M_{Z'}$& $3.5\,{\rm TeV}<M_{Z'}<220\,{\rm TeV}$ & $3.5\,{\rm TeV}<M_{Z'}<56\,{\rm TeV}$ & $3.5\,{\rm TeV}<M_{Z'}<7.4\,{\rm TeV}$ \\
$g_{B-L}$& $0.0175< g_{B-L}<1.1$ & $0.15< g_{B-L}<1.1$ & $0.5< g_{B-L}<1.1$ \\
$\lambda_{\phi}$& $5\times 10^{-4}<\lambda_{\phi}$ & $0.04<\lambda_{\phi}$ & $0.4<\lambda_{\phi}$ \\
\hline
\end{tabular}
\caption{
  \label{tab:3}
\it \small The range of coupling constants corresponding to benchmark points in Table~\ref{tab:2}.
}

\end{table}

Tables~\ref{tab:2} and \ref{tab:3} 
detail three new benchmark points (Scenarios  5, 6 $\&$ 7) where lepton flavour asymmetry is generated in the classically 
scale-invariant $B-L$ model with
the Majorana masses in the GeV range. In these scenarios we 
successively lower the vev of the Coleman-Weinberg filed $\langle |\phi |\rangle$ from $10^5$ to $3.4 \times 10^3$ GeV. The second column,
Scenario 6 gives the values of the portal coupling $\lambda_{\rm P} \simeq 10^{-5}$ GeV which is in agreement
with the presently available Higgs data 
constraints and can be probed by the future experiments\cite{Englert:2013gz}.

The third column (Scenario 7) in the Tables~\ref{tab:2} and \ref{tab:3} enters the regime where $\lambda_{\rm P}$ approaches $10^{-3}$. 
(To achieve this we brought the three Majorana masses closer together relative to Scenarios 5 and 6.)
In this case, the Higgs potential is automatically stabilised by the positive contribution $\propto \lambda_{\rm P}^2$ 
to the Higgs self-coupling $\lambda_H$ beta function~ \cite{Lykken,Hambye:2013dgv}. 

We also show the values of the Majorana $Y^{\rm M}$ and the average value of the Dirac Yukawa 
$\left<Y^{\rm D}\right>$ couplings\footnote{The latter is computed as the average of $\sqrt 2 m M/v$.} along with the ranges for
$g_{B-L}$, or equivalently, the $Z'$ vector boson mass,
and the self-coupling of the CW scalar. The lower bound on the $Z'$ mass in the Table is the experimental bound
$M_{Z'} \ge 3.5$ TeV which is then translated into the lower bounds on $g_{B-L}$ via
\begin{equation}
\label{eqn:MZ2}
m_{Z'}\,=\,Q_\phi\cdot g_{B-L}\langle |\phi |\rangle
\,=\,2\, g_{B-L}\langle |\phi |\rangle
\,.
\end{equation}
The upper bounds on  $M_{Z'}$ in the Table follow from the requirement of perturbativity in the coupling $\alpha_{B-L} \le 0.1$, 
which gives $g_{B-L} \lesssim 1.1$.
For the CW self-coupling $\lambda_\phi$ the lower bounds are determined via ({\it c.f.} Eq.~\eqref{eq:rad4}),
\begin{equation}
  \label{eq:rad4222}
  \lambda_{\phi}\,=\, \frac{33}{8\pi^2} (Q_\phi \cdot g_{B-L})^4 
  \,=\, \frac{33}{8\pi^2} (2\, g_{B-L})^4 
  \,.
\end{equation}

\section{Conclusions}
\label{sec:five}

In a theory with no input mass scales, the
Coleman-Weinberg mechanism generates a symmetry-breaking vev and
the mass for the associated scalar from radiative corrections. These
scales are natural in the sense that they are automatically
exponentially suppressed compared to the UV scale at which we
initialise the theory. The portal interaction between the Coleman-Weinberg scalar and the Higgs then generates the 
electroweak symmetry breaking scale. 

\medskip
We have shown that this theory for a wide range of parameters can generate the observed 
value of matter-anti-matter asymmetry via the leptogenesis mechanism due to Majorana neutrino oscillations.

\medskip
As a bonus the $B-L$ theory considered in this paper 
automatically contains sterile Majorana neutrinos with masses in the window roughly between
200 MeV and 500 GeV, and a heavier $Z'$ boson. 
The presently available Higgs data provide valuable
constraints on the parameter space of the model, and future experimental data on
Higgs decays  will further constrain
model parameters in the Higgs sector~\cite{Englert:2013gz}. Additional experimental constraints will come from searches of the sub-TeV-scale
sterile neutrinos via a combination of the double beta decay, electroweak precision data, the LHC searches and the high intensity frontier,
see  e.g. \cite{Drewes:2013gca,Boyarsky:2012rt} for recent reviews.
The third ingredient comes from the searches of $Z'$ vector bosons, see e.g. \cite{Basso:2008iv}.
All these should ultimately provide the discovery potential for this classically conformal theory.

\medskip
The follow-up paper~\cite{Khoze:2013uia} presents an implementation of the slow-roll inflation mechanism in a BSM theory with classical scale invariance.
This is achieved by
introducing an additional singlet scalar field to the portal interactions of the theory 
and requiring that this singlet is also non-minimally coupled to gravity. 
At the same time, the same singlet provides a viable dark matter candidate in this theory. 
Furthermore, the SM Higgs potential is stabilised by the Higgs portal interactions with the Coleman-Weinberg scalar.

\medskip
These results support the BSM model-building strategy
which is based on classically scale-invariant extensions 
of the SM with portal-type interactions involving the Higgs field as well as other microscopic scalars. These theories appear to be 
capable of addressing all the shortcomings of the SM mentioned in the Introduction.

\medskip
\section*{Acknowledgements}
We would like to thank Steven Abel, Christoph Englert, Joerg Jaeckel and Michael Spannowsky for stimulating
discussions and collaboration on related topics.
VVK is supported in part by STFC through the IPPP grant ST/G000905/1 and in part by the Wolfson Foundation and Royal Society. 
GR acknowledges the receipt of a Durham Doctoral Studentship.
%

\bigskip

\bibliographystyle{h-physrev5}

\begin{thebibliography}{99}

\bibitem{orig} 
  F.~Englert and R.~Brout,
  ``Broken Symmetry and the Mass of Gauge Vector Mesons,''
  Phys.\ Rev.\ Lett.\  {\bf 13} (1964) 321,\\
  P.~W.~Higgs,
  ``Broken symmetries, massless particles and gauge fields,''
  Phys.\ Lett.\  {\bf 12} (1964) 132 and
  ``Broken Symmetries and the Masses of Gauge Bosons,''
  Phys.\ Rev.\ Lett.\  {\bf 13} (1964) 508,\\
  G.~S.~Guralnik, C.~R.~Hagen and T.~W.~B.~Kibble,
  ``Global Conservation Laws and Massless Particles,''
  Phys.\ Rev.\ Lett.\  {\bf 13} (1964) 585.

\bibitem{ATLAS:2012gk}
  G.~Aad {\it et al.}  [ATLAS Collaboration],
  ``Observation of a new particle in the search for the Standard Model Higgs boson with the ATLAS detector at the LHC,''
  Phys.\ Lett.\ B {\bf 716} (2012) 1,
  arXiv:1207.7214 [hep-ex].

\bibitem{CMS:2012gu}
  S.~Chatrchyan {\it et al.}  [CMS Collaboration],
  ``Observation of a new boson at a mass of 125 GeV with the CMS experiment at the LHC,''
  Phys.\ Lett.\ B {\bf 716} (2012) 30,
  arXiv:1207.7235 [hep-ex].
  
\bibitem{Coleman:1973jx}
  S.~R.~Coleman and E.~J.~Weinberg,
  ``Radiative Corrections as the Origin of Spontaneous Symmetry Breaking,''
  Phys.\ Rev.\ D {\bf 7} (1973) 1888.
\bibitem{Bardeen:1995kv}
  W.~A.~Bardeen,
  ``On naturalness in the standard model,''
  FERMILAB-CONF-95-391-T.

\bibitem{Meissner:2006zh}
  K.~A.~Meissner and H.~Nicolai,
  ``Conformal Symmetry and the Standard Model,''
  Phys.\ Lett.\ B {\bf 648} (2007) 312,
  hep-th/0612165.
  
\bibitem{Hempfling:1996ht}
  R.~Hempfling,
  ``The Next-to-minimal Coleman-Weinberg model,''
  Phys.\ Lett.\ B {\bf 379} (1996) 153,
  hep-ph/9604278.

  
\bibitem{Chang:2007ki}
  W.~-F.~Chang, J.~N.~Ng and J.~M.~S.~Wu,
  ``Shadow Higgs from a scale-invariant hidden U(1)(s) model,''
  Phys.\ Rev.\ D {\bf 75} (2007) 115016,
  hep-ph/0701254.

\bibitem{Foot:2007as}
  R.~Foot, A.~Kobakhidze and R.~R.~Volkas,
  ``Electroweak Higgs as a pseudo-Goldstone boson of broken scale invariance,''
  Phys.\ Lett.\ B {\bf 655} (2007) 156,
  arXiv:0704.1165 [hep-ph],\\
  ``Stable mass hierarchies and dark matter from hidden sectors in the scale-invariant standard model,''
  Phys.\ Rev.\ D {\bf 82} (2010) 035005,
  arXiv:1006.0131 [hep-ph].
  
\bibitem{Iso:2009ss}
  S.~Iso, N.~Okada and Y.~Orikasa,
  ``Classically conformal $B^-$ L extended Standard Model,''
  Phys.\ Lett.\ B {\bf 676} (2009) 81,
  arXiv:0902.4050 [hep-ph].
 
\bibitem{Iso:2012jn}
  S.~Iso and Y.~Orikasa,
  ``TeV Scale B-L model with a flat Higgs potential at the Planck scale - in view of the hierarchy problem -,''
  arXiv:1210.2848 [hep-ph].
  
\bibitem{Englert:2013gz}
  C.~Englert, J.~Jaeckel, V.~V.~Khoze and M.~Spannowsky,
  ``Emergence of the Electroweak Scale through the Higgs Portal,''
  JHEP {\bf 1304} (2013) 060,
  arXiv:1301.4224 [hep-ph].

\bibitem{Chun:2013soa}
  E.~J.~Chun, S.~Jung and H.~M.~Lee,
  ``Radiative generation of the Higgs potential,''
  Phys.\ Lett.\ B {\bf 725} (2013) 158,
  arXiv:1304.5815 [hep-ph].
  
\bibitem{Heikinheimo:2013fta}
  M.~Heikinheimo, A.~Racioppi, M.~Raidal, C.~Spethmann and K.~Tuominen,
  ``Physical Naturalness and Dynamical Breaking of Classical Scale Invariance,''
  arXiv:1304.7006 [hep-ph].
  
\bibitem{Hambye:2013dgv}
  T.~Hambye and A.~Strumia,
  ``Dynamical generation of the weak and Dark Matter scale,''
  arXiv:1306.2329 [hep-ph].

 
 \bibitem{Higgs.portal}
 T.~Binoth and J.~J.~van der Bij,
  ``Influence of strongly coupled, hidden scalars on Higgs signals,''
  Z.\ Phys.\  C {\bf 75}, 17 (1997),
 arXiv:hep-ph/9608245,\\
  R.~Schabinger and J.~D.~Wells,
  ``A Minimal spontaneously broken hidden sector and its impact on Higgs boson physics at the large hadron collider,''
  Phys.\ Rev.\ D {\bf 72} (2005) 093007,  hep-ph/0509209,\\
  B.~Patt and F.~Wilczek,
  ``Higgs-field portal into hidden sectors,''
  arXiv:hep-ph/0605188,\\
  C.~Englert, T.~Plehn, D.~Zerwas and P.~M.~Zerwas,
  ``Exploring the Higgs portal,''
  Phys.\ Lett.\ B {\bf 703} (2011) 298, arXiv:1106.3097 [hep-ph].

\bibitem{Fukugita:1986hr}
  M.~Fukugita and T.~Yanagida,
  ``Baryogenesis Without Grand Unification,''
  Phys.\ Lett.\ B {\bf 174} (1986) 45.
  
\bibitem{Manton:1983nd}
  N.~S.~Manton,
  ``Topology in the Weinberg-Salam Theory,''
  Phys.\ Rev.\ D {\bf 28} (1983) 2019.

\bibitem{Kuzmin:1985mm}
  V.~A.~Kuzmin, V.~A.~Rubakov and M.~E.~Shaposhnikov,
  ``On the Anomalous Electroweak Baryon Number Nonconservation in the Early Universe,''
  Phys.\ Lett.\ B {\bf 155} (1985) 36.

  
\bibitem{Davidson:2002qv}
  S.~Davidson and A.~Ibarra,
  ``A Lower bound on the right-handed neutrino mass from leptogenesis,''
  Phys.\ Lett.\ B {\bf 535} (2002) 25,
  hep-ph/0202239.

\bibitem{Davidson:2008bu}
  S.~Davidson, E.~Nardi and Y.~Nir,
  ``Leptogenesis,''
  Phys.\ Rept.\  {\bf 466} (2008) 105,
  arXiv:0802.2962 [hep-ph].
  
\bibitem{Akhmedov:1998qx}
  E.~K.~Akhmedov, V.~A.~Rubakov and A.~Y.~.Smirnov,
  ``Baryogenesis via neutrino oscillations,''
  Phys.\ Rev.\ Lett.\  {\bf 81} (1998) 1359,
  hep-ph/9803255.
  
\bibitem{Asaka:2005pn}
  T.~Asaka and M.~Shaposhnikov,
  ``The nuMSM, dark matter and baryon asymmetry of the universe,''
  Phys.\ Lett.\ B {\bf 620} (2005) 17,
  hep-ph/0505013.
  
\bibitem{Drewes:2012ma}
  M.~Drewes and B.~Garbrecht,
  ``Leptogenesis from a GeV Seesaw without Mass Degeneracy,''
  JHEP {\bf 1303} (2013) 096,
  arXiv:1206.5537 [hep-ph].
   
\bibitem{Lykken}
  J.~Lykken,  ``Higgs without Supersymmetry,''
  Talk at the MITP Workshop, Mainz, Germany, March 18-22, 2013.
  
\bibitem{EliasMiro:2012ay}
  J.~Elias-Miro, J.~R.~Espinosa, G.~F.~Giudice, H.~M.~Lee and A.~Strumia,
  ``Stabilization of the Electroweak Vacuum by a Scalar Threshold Effect,''
  JHEP {\bf 1206} (2012) 031,
  arXiv:1203.0237 [hep-ph].

\bibitem{Mohapatra:1980qe}
  R.~N.~Mohapatra and R.~E.~Marshak,
  ``Local B-L Symmetry of Electroweak Interactions, Majorana Neutrinos and Neutrono Oscillations,''
  Phys.\ Rev.\ Lett.\  {\bf 44} (1980) 1316
   [Erratum-ibid.\  {\bf 44} (1980) 1643];\\
  R.~E.~Marshak and R.~N.~Mohapatra,
  ``Quark - Lepton Symmetry and B-L as the U(1) Generator of the Electroweak Symmetry Group,''
  Phys.\ Lett.\ B {\bf 91}, 222 (1980);\\
  C.~Wetterich,
  ``Neutrino Masses and the Scale of B-L Violation,''
  Nucl.\ Phys.\ B {\bf 187} (1981) 343,\\
  S.~Khalil,
  ``Low scale B - L extension of the Standard Model at the LHC,''
  J.\ Phys.\ G {\bf 35} (2008) 055001
  hep-ph/0611205,\\
  S.~Khalil and A.~Masiero,
  ``Radiative B-L symmetry breaking in supersymmetric models,''
  Phys.\ Lett.\ B {\bf 665} (2008) 374,
  arXiv:0710.3525 [hep-ph],\\
  
\bibitem{Basso:2008iv}
  L.~Basso, A.~Belyaev, S.~Moretti and C.~H.~Shepherd-Themistocleous,
  ``Phenomenology of the minimal B-L extension of the Standard model: Z' and neutrinos,''
  Phys.\ Rev.\ D {\bf 80} (2009) 055030,
  arXiv:0812.4313 [hep-ph],\\
  L.~Basso, A.~Belyaev, S.~Moretti, G.~M.~Pruna and C.~H.~Shepherd-Themistocleous,
  ``$Z'$ discovery potential at the LHC in the minimal $B-L$ extension of the Standard Model,''
  Eur.\ Phys.\ J.\ C {\bf 71} (2011) 1613,
  arXiv:1002.3586 [hep-ph],\\
  L.~Basso, S.~Moretti and G.~M.~Pruna,
  ``A Renormalisation Group Equation Study of the Scalar Sector of the Minimal B-L Extension of the Standard Model,''
  Phys.\ Rev.\ D {\bf 82} (2010) 055018,
  arXiv:1004.3039 [hep-ph].
    
\bibitem{Barbieri:1999ma}
  R.~Barbieri, P.~Creminelli, A.~Strumia and N.~Tetradis,
  ``Baryogenesis through leptogenesis,''
  Nucl.\ Phys.\ B {\bf 575} (2000) 61,
  hep-ph/9911315,\\
  A.~Abada, S.~Davidson, F.~-X.~Josse-Michaux, M.~Losada and A.~Riotto,
  ``Flavor issues in leptogenesis,''
  JCAP {\bf 0604} (2006) 004,
  hep-ph/0601083,\\
  E.~Nardi, Y.~Nir, E.~Roulet and J.~Racker,
  ``The Importance of flavor in leptogenesis,''
  JHEP {\bf 0601} (2006) 164,
  hep-ph/0601084,\\
  S.~Blanchet and P.~Di Bari,
  JCAP {\bf 0703} (2007) 018,
  hep-ph/0607330.

\bibitem{Pilaftsis:2003gt}
  A.~Pilaftsis and T.~E.~J.~Underwood,
  ``Resonant leptogenesis,''
  Nucl.\ Phys.\ B {\bf 692} (2004) 303,
  hep-ph/0309342 and 
  ``Electroweak-scale resonant leptogenesis,''
  Phys.\ Rev.\ D {\bf 72} (2005) 113001,
  hep-ph/0506107.
  
\bibitem{Sigl:1992fn}
  G.~Sigl and G.~Raffelt,
  ``General kinetic description of relativistic mixed neutrinos,''
  Nucl.\ Phys.\ B {\bf 406} (1993) 423.
  
\bibitem{Besak:2012qm}
  D.~Besak and D.~Bodeker,
  ``Thermal production of ultrarelativistic right-handed neutrinos: Complete leading-order results,''
  JCAP {\bf 1203} (2012) 029,
  arXiv:1202.1288 [hep-ph].
  
\bibitem{Schwinger:1960qe}
  J.~S.~Schwinger,
  ``Brownian motion of a quantum oscillator,''
  J.\ Math.\ Phys.\  {\bf 2} (1961) 407.
  
\bibitem{Keldysh:1964ud}
  L.~V.~Keldysh,
  ``Diagram technique for nonequilibrium processes,''
  Zh.\ Eksp.\ Teor.\ Fiz.\  {\bf 47} (1964) 1515
   [Sov.\ Phys.\ JETP {\bf 20} (1965) 1018].

\bibitem{Garbrecht:2011aw}
  B.~Garbrecht and M.~Herranen,
  ``Effective Theory of Resonant Leptogenesis in the Closed-Time-Path Approach,''
  Nucl.\ Phys.\ B {\bf 861} (2012) 17,
  arXiv:1112.5954 [hep-ph].
  
\bibitem{Garny:2011hg}
  M.~Garny, A.~Kartavtsev and A.~Hohenegger,
  ``Leptogenesis from first principles in the resonant regime,''
  Annals Phys.\  {\bf 328} (2013) 26,
  arXiv:1112.6428 [hep-ph].
  
\bibitem{Casas:2001sr} 
  J.~A.~Casas and A.~Ibarra,
  ``Oscillating neutrinos and muon ---> e, gamma,''
  Nucl.\ Phys.\ B {\bf 618}, 171 (2001),
  hep-ph/0103065.
 
 
 \bibitem{Drewes:2013gca}
  M.~Drewes,
  ``The Phenomenology of Right Handed Neutrinos,''
  arXiv:1303.6912 [hep-ph].

\bibitem{Boyarsky:2012rt}
  A.~Boyarsky, D.~Iakubovskyi and O.~Ruchayskiy,
  ``Next decade of sterile neutrino studies,''
  Phys.\ Dark Univ.\  {\bf 1} (2012) 136,
  arXiv:1306.4954 [astro-ph.CO].

\bibitem{Khoze:2013uia}
  V.~V.~Khoze,
  ``Inflation and Dark Matter in the Higgs Portal of Classically Scale Invariant Standard Model,''
  arXiv:1308.6338 [hep-ph].
 
  

\end{thebibliography}

\end{document}